\documentclass[twocolumn,showpacs,preprintnumbers]{revtex4}
\def\pref#1{(\ref{#1})}
\usepackage{graphicx} 
\usepackage{amssymb}
\usepackage{amsmath}
\usepackage{hyperref}

\begin{document}


\title{Intersubband Polaritons in the Electrical Dipole Gauge}

\date{\today}

\author{Yanko Todorov}
\author{Carlo Sirtori}
\affiliation{Laboratoire "Mat{\'e}riaux et Ph{\'e}nom{\`{e}}nes
Quantiques", Universit{\'e} Paris Diderot-Paris 7, CNRS- UMR 7162,
75013 Paris, France}

\begin{abstract}
We provide a theoretical description for the coupling between the
intersubband excitations of a bi-dimensional electron gas with the electromagnetic field.
This description, based on the electrical dipole gauge, applies to an arbitrary
quantum heterostructure embedded in a general multilayered
waveguide or a microcavity.  We show that the dipole gauge Hamiltonian
automatically takes into account the Coulomb interactions in this system.
Furthermore, it can be conveniently expressed in terms of the many-body collective plasmon modes,
which interact both with each other and with the light field. The dipole gauge therefore provides
a suitable framework for the study of solid state
Quantum Electrodynamics (QED) phenomena, such as the ultra-strong light-matter interaction regime,
occurring at very high electronic densities.
\end{abstract}
\maketitle

\section{Introduction}

In the description of a physical problem dealing with light-matter
interaction there is always a degree of freedom in the choice of
the potentials associated with the electromagnetic field. Even
though the physical phenomena are obviously independent from the
particular vector or scalar potentials used, the choice of the
gauge is a critical issue, as it can be more or less adequate for
the physical interpretation of the phenomena. In the literature
there is already a consensus that the interaction of
non-relativistic bound charges is very conveniently expressed in
terms of the dipole gauge \cite{Babiker_Loudon_1983}. This has
been firstly identified by Power and Zienau in the 50's
\cite{Power_Zienau_1957} and subsequently by Woolley
\cite{Woolley_1971}. In this formulation, called
Power-Zienau-Woolley (PZW) gauge, the sources are electric and
magnetic polarization fields and the coupling occurs via the
intensities of the displacement field $\mathbf{D}$ and the
magnetic field $\mathbf{H}$ rather than using the scalar and
vector potentials $V$ and $\mathbf{A}$. In the case where the
magnetic interactions in the system can be neglected, the PZW
gauge is also called dipole gauge since the interaction
Hamiltonian contains only the coupling between the material
polarization $\mathbf{P}$ with the light field.

In our article we propose to apply the dipole gauge to the case of
intersubband transitions interacting with a photonic cavity mode.
The motivation of our approach stems from the fact that we are interested to investigate a regime, called “ultra-strong coupling” firstly introduced by Ciuti et al. \cite{Ciuti_PhysRevB_2005}. This regime of light-matter interaction is 
attainable in a bi-dimensional electron gas with very high density embedded 
into a photonic microcavity \cite{Dini_2003, Ciuti_PhysRevB_2005, Todorov_PRL2010} 
and is characterized by the fact that the coupling and the material excitation energies are comparable quantities. However, the very high electronic densities increase on one hand the light-matter interaction and on the other renormalise the 
transition energies of the system due to a collective effect, the "depolarization
shift" \cite{Ando_Fowler_Stern_1982}. This effect cannot be neglected in 
the limit of ultra-strong coupling, however it was not explicitly considered in 
the initial study described in Ref. \cite{Ciuti_PhysRevB_2005}. In our work we show that the dipole gauge Hamiltonian handles the light-matter interaction and the depolarization effect on the same footing. The underlying physical picture is that 
the active polarization that couples to the photonic mode is not a single particle electronic transition between confined states but rather a collective electronic 
mode, - a plasmonic mode -, arising from electrons distributed in different subbands, yet phased by the Coulomb interaction. The latter point is one of the main conclusions of our theoretical investigation.

The correct description of the collective electronic excitations
and their interaction with light can be obtained in the Coulomb
gauge, only if both the vector potential $\mathbf{A}$ and scalar
potential $V$ are considered. This has already been noticed in
studies of ensembles of two level systems interacting with light
\cite{Keeling_2007}. In the absence of the cavity, thus in the weak coupling 
limit, collective effects can be identified with the scalar potential $V$ \cite{Nikonov_1997}, which describes an instantaneous interaction between electrons \cite{Book_Cohen_Ph_At}. However, when dealing with the resonant coupling of a microcavity with an electronic transition the observable quantity is the retarded electromagnetic field that can be obtained through a proper combination of both $V$ and $\mathbf{A}$ \cite{Book_Cohen_Ph_At}. In order to have a general Hamiltonian, 
which is valid from the weak to the ultra-strong coupling regimes, it 
is therefore essential to include the Coulomb potential $V$, thus 
completing the study of Ref. \cite{Ciuti_PhysRevB_2005}. Such Hamiltonian is 
readily obtained in the dipole gauge, where the matter degrees of 
freedom are gathered in the polarization field $\mathbf{P}$, which not only describes the interactions between the electrons, but also couples to a material independent and retarded photonic field described by the electric displacement $\mathbf{D}$ \cite{Babiker_Loudon_1983, Book_Cohen_Ph_At}. Indeed, as we previously pointed out in Ref. \cite{Todorov_PRL2010}, the depolarization effect in the bi-dimensional electron gas is contained in the quadratic $\mathbf{P}^2$ term of the dipolar Hamiltonian. As a result, our model correctly describes the local field effects arising from the very different spatial scales between the electronic and photonic confinement. The results of Ref. \cite{Ciuti_PhysRevB_2005} are contained in our formalism and recovered only in the opposite limit, i.e. when the electronic polarization fills the whole cavity volume. The importance of the spatial overlap factor was also outlined in our experimental study of the ultra-strong coupling regime \cite{Todorov_PRL2010, PJouy_2011} as it allows to correlate both polaritonic and weak-coupling absorption data. On a more fundamental level, we show how the spatial confinement of the electronic polarization leads to the No-go theorem for intersubband transitions 
\cite{Birula_1979, Nataf_Ciuti_2010}.

Our paper is organized as follows. In part \ref{Part_I} we
establish the Hamiltonian of the system in the dipole gauge.
We consider a very general case of an arbitrary
heterostructure embedded into a general planar waveguide
multilayer. The microscopic expression of the intersubband
polarization field is derived in section \ref{Sec_MicroDefP},
in the long-wavelength approximation. This microscopic expression 
relies directly on the electronic wavefunctions, and therefore 
allows to go beyond the semi-classical model employed in 
Ref.\cite{Todorov_PRL2010}, as it applies to an arbitrary heterostructure 
potential, with an arbitrary number of occupied subbands.
  
In part \ref{Part_II} we adapt the general dipolar Hamiltonian
obtained in part \ref{Part_I} to the case of a series of highly
doped quantum wells and we express it in terms of the collective
electronic excitations of the system. We also provide a version of
this "plasma Hamiltonian" for the case of 0D resonators (section
\ref{Sec_0D}). In section \ref{min_coupling_gauge} we study
the correspondence between the plasma Hamiltonian, truncated for the case
of a single intersubband transition and a single waveguide mode,
and the corresponding Hamiltonian in the Coulomb gauge. This allows
us to connect our formalism with previous work \cite{Ciuti_PhysRevB_2005}. 
In particular, we show that the dipole gauge provides automatically the relevant
contributions due to the Coulomb interaction of the system, that were missing in 
Ref. \cite{Ciuti_PhysRevB_2005}.

The formalism is applied to study the properties of the polariton
states in part \ref{Part_III}. In section \ref{Sec_Dispersion} we
examine the polariton dispersion, and in section \ref{Sec_NoGo} we
discuss the No-go theorem for intersubband transitions. We show in
section \ref{DielFuncSection} that our formalism is consistent
with the effective medium approach, in full agreement with what
has been proposed in the literature \cite{Zaluzny_Nalewajko_1999}.
Most of the technical details have been gathered in the
appendixes.

\section{Interaction Hamiltonian}\label{Part_I}

\subsection{General considerations}\label{Sec_Gen_Cons}

The interaction between light and quantum heterostructures is
usually studied in planar multilayered systems. This geometry is
most naturally compatible with the epitaxial growth. A very
general system is described in Figure \ref{Fig1}(a). It
consists of homogeneous, non-absorbing and non-dispersive
dielectric layers described by real dielectric constants
$\varepsilon_i$, embedded between two infinite semiplanes that act
as optical claddings. We can then define a piecewise dielectric
function $\varepsilon(z)$, with $\varepsilon(z)=\varepsilon_i$ in
the $i^{\mathrm{th}}$ layer, $z$ being the growth axis. The
multi-layered dielectric stack defines guided modes that confine
the light field around the heterostructures. Note that, at this
point, $\varepsilon(z)$ does not include the resonant contribution
from the electronic transition in the heterostructres. This
contribution will be included through the coupling between the
guided modes and the electronic polarization. The full Hamiltonian
of the system writes:

\begin{equation}\label{Eq1}
\hat{H}= \hat{H}_e+\hat{H}_p+\hat{H}_{\mathrm{int}}
\end{equation}

Here $\hat{H}_e$ is the Hamiltonian of the electron gas in the
heterostructures, which is provided explicitly in section
\ref{Sec_H_e}, and $\hat{H}_p$ is the photon Hamiltonian:

\begin{equation}\label{H_p0}
\hat{H}_p =\int \Big{[} \frac{1}{2\varepsilon_0 \varepsilon(z)
}\mathbf{\hat{D}}^2(\mathbf{r}) +
\frac{\mu_0}{2}\mathbf{\hat{H}}^2(\mathbf{r})\Big{]} d^3
\mathbf{r}
\end{equation}

with $\mathbf{\hat{D}}(\mathbf{r})$ and
$\mathbf{\hat{H}}(\mathbf{r})$ respectively the displacement field
and the magnetic field operators. The interaction Hamiltonian
$\hat{H}_{\mathrm{int}}$ in the electrical dipole gauge is written
as, neglecting the magnetic interactions \cite{Book_Welsch}:

\begin{equation}\label{H_int}
\hat{H}_{\mathrm{int}} = \int \frac{1}{\varepsilon_0
\varepsilon(z)} \Big{[}-\mathbf{\hat{D}}(\mathbf{r})\cdot
\mathbf{\hat{P}}(\mathbf{r}) + \frac{1}{2}
 \mathbf{\hat{P}}^2(\mathbf{r})\Big{]} d^3 \mathbf{r}
\end{equation}

Here $\mathbf{\hat{P}}(\mathbf{r})$ is the polarization density
operator of the electron gas. This is a central quantity in our
theory, and it is provided explicitly in section
\ref{Sec_MicroDefP} from a microscopic model. One issue that we
will discuss in details in the article is the role of the
quadratic interaction term, $\mathbf{\hat{P}}^2(\mathbf{r})$, which describes
the self-interaction of the electronic polarization, and therefore contains
the effects of the dipole-dipole interactions.

Most generally, the expression of the interaction \pref{H_int}
should be written as a non-local expression, which takes into
account the spatial dispersion of the electromagnetic response of
the medium \cite{Suttorp_2007}. However, as illustrated in Figure
\ref{Fig1}(a), in the systems that we study, the
heterostructures embedded in the multilayered stack have a typical
extension in the growth axis (the $z$-axis) that is much smaller
than the wavelength. This justifies the local form of the
interaction postulated in \pref{H_int}.

We can evaluate the convenience of the dipole gauge already at the
classical level, where the displacement field
$\mathbf{D}(\mathbf{r})$ is determined only by the charges
exterior to the system. On a quantum mechanical level, this means
that $\mathbf{\hat{D}}(\mathbf{r})$ describes a purely transverse
field \cite{Book_Cohen_Ph_At}. Therefore, neglecting the
dissipation in the system, we can consider
$\mathbf{\hat{D}}(\mathbf{r})$ as a free photon field, independent
from the electronic polarization of the heterostructures. This can
be clearly observed, for instance, in the normal component of the
displacement field $\mathbf{\hat{D}}_z$, that, while it couples to
the intersubband polarization, is continuous across the
interfaces. We therefore consider the derivation of
$\mathbf{\hat{D}}(\mathbf{r})$ as a separate problem, which is
dependent only on the particular arrangement of the multilayered
stack. In this approach the resonant contribution of the
electronic intersubband transitions taking place in the active
media are contained separately in the polarization operator
density $\mathbf{\hat{P}}$, and are active only through the
interaction term $\hat{H}_\mathrm{int}$. Later on (section
\ref{DielFuncSection}) we shall see that this approach leads to an
effective medium treatment of the system.

In our treatment the dissipation effects will be neglected. The
latter can be taken into account as in the classical paper by
Huttner and Barnet \cite{Huttner_Barnet_1992}, by including a
dissipative bath with continuous degrees of freedom. This approach
will however require more precise definition of the displacement
field $\mathbf{\hat{D}}$, as it leads to a noise contribution
from the dissipative bath \cite{Huttner_Barnet_1992, Book_Welsch}.
Here we will restrict to a fully Hamiltonian treatment. The
dissipation will be taken into account only in a phenomenological
way in section \ref{DielFuncSection}, by adding a small imaginary
part to the eigenfrequencies of the system
\cite{Dutra_Furuya_1998}.

\subsection{Free photon Hamiltonian}\label{Sec_FreePhot}

\begin{figure}
\includegraphics[scale=0.15]{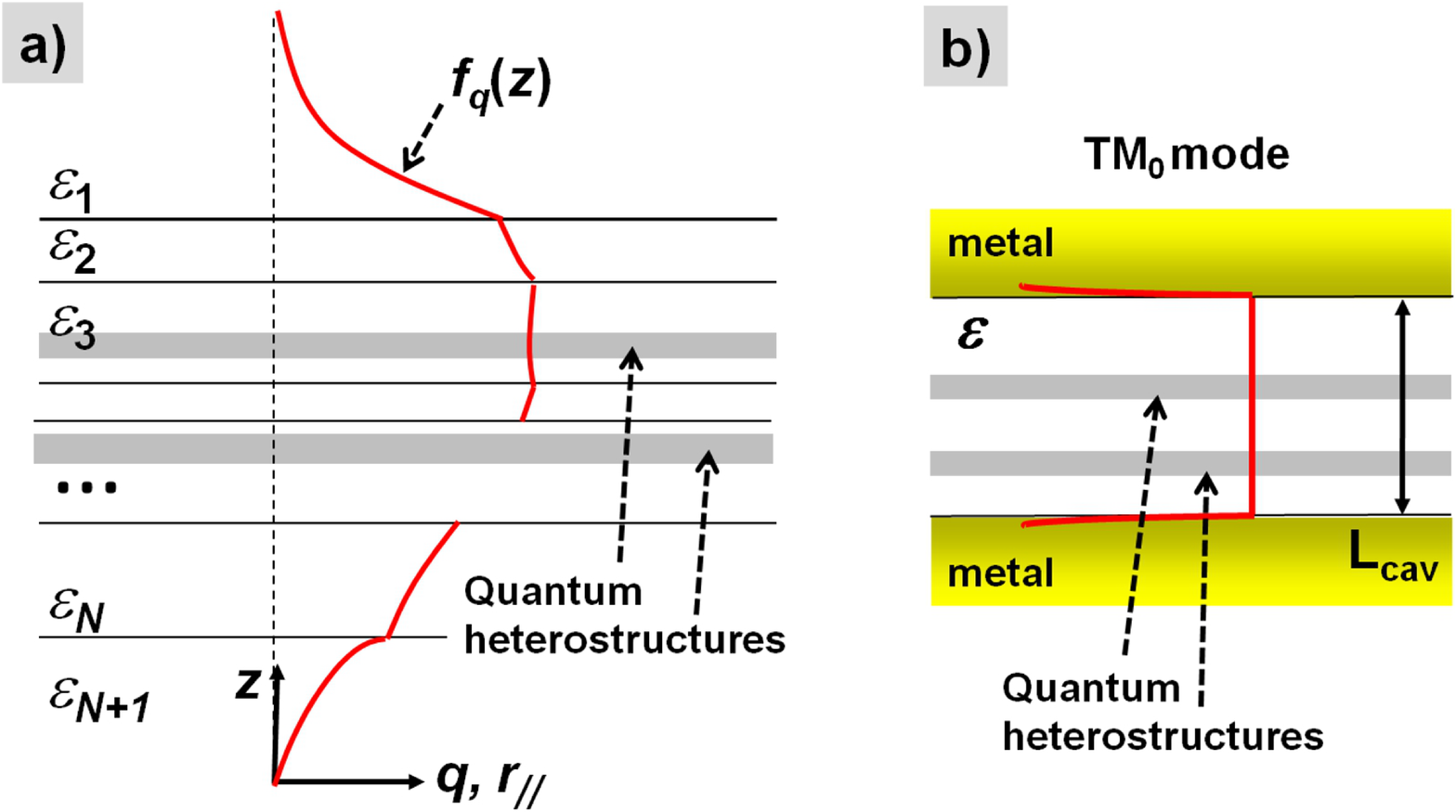}
\caption {a) General planar multi-layered system with a piecewise
dielectric function $\varepsilon(z)$, supporting guided modes.
There are quantum heterostructures embedded inside the
multi-layered stack. b) $\mathrm{TM}_0$ mode guided between two
metallic plates.} \label{Fig1}
\end{figure}

For the quantum description of electromagnetic field, we use the
basis of the guided modes of the multilayered stack, that are
bounded in space. Due to the translational invariance of the
system in the plane perpendicular to the $z$-axis, the guided
modes are characterized by their in-plane wavevector $\mathbf{q}$
and their energy $\hbar \omega_{c \mathbf{q}}$. The function
$\omega_{c \mathbf{q}}=\omega_{c \mathbf{q}}(|\mathbf{q}|)$
defines the dispersion relation of the guided modes. The
heterostructures interact only with the TM-polarized modes, in
order to respect the selection rule of intersubband transitions
\cite{Book_Helm}. Using the general quantization rules
\cite{Book_Cohen_Ph_At, Mahan_2010}, we can assign bosonic
creation and annihilation operators $a_\mathbf{q}^\dagger$ and
$a_\mathbf{q}$ to each guided mode. The components of the
quantized electromagnetic TM free field are then:

\begin{eqnarray}
\mathbf{\hat{H}} = \sum_\mathbf{q}
iA_\mathbf{q}(\mathbf{e}_q\wedge \mathbf{e}_z)
f_\mathbf{q}(z)e^{i\mathbf{q}\mathbf{r}_{\parallel}}(a_\mathbf{q}+a_{-\mathbf{q}}^\dagger)
\\ \label{D_z_general}
\mathbf{\hat{D}}_z = \sum_\mathbf{q} iA_\mathbf{q}\mathbf{e}_z
\frac{|\mathbf{q}|}{\omega_{c \mathbf{q}}}
f_\mathbf{q}(z)e^{i\mathbf{q}\mathbf{r}_{\parallel}}(a_\mathbf{q}-a_{-\mathbf{q}}^\dagger)
\\
\mathbf{\hat{D}}_{\parallel}=-\sum_\mathbf{q}
A_\mathbf{q}\mathbf{e}_\mathbf{q} \frac{1}{\omega_{c
\mathbf{q}}}\frac{\mathrm{d}f_\mathbf{q}(z)}{\mathrm{d}z}
e^{i\mathbf{q}\mathbf{r}_{\parallel}}(a_\mathbf{q}-a_{-\mathbf{q}}^\dagger)
\\ \label{A_q}
A_\mathbf{q} = \Big{(}\frac{\hbar \omega_{c \mathbf{q}} }{2\mu_0 S
L_\mathbf{q}}\Big{)}^{1/2}
\end{eqnarray}

Here $S$ is the area of the system, and $\mathbf{r}_{\parallel}$
is the in-plane position vector. We have introduced the unit
vectors $\mathbf{e}_z$ and $\mathbf{e}_\mathbf{q}
=\mathbf{q}/|\mathbf{q}|$, and the symbol "$\wedge$" designs the
vector product. The constant $A_\mathbf{q}$ is the vacuum field
intensity of the guided modes. It has been derived, for this
particular system, in Appendix \ref{AppNormEMfield}. The function
$f_\mathbf{q}(z)$ describes the lateral profile of the guided
modes, and $L_\mathbf{q}$ is a normalization coefficient:

\begin{equation}
\int_{-\infty}^{+\infty} f_\mathbf{q}(z)^2 dz =L_\mathbf{q}
\end{equation}

If we set the maximum of the dimensionless function $f_\mathbf{q}(z)$ equal to $1$,
with this definition $L_\mathbf{q}$ can be regarded as the
effective thickness of the multilayer system. The function
$f_\mathbf{q}(z)$ satisfies the Helmholtz equation:

\begin{equation}\label{f_q_equation}
\frac{\mathrm{d}^2f_\mathbf{q}(z)}{\mathrm{d}z^2} -
\mathbf{q}^2f_\mathbf{q}(z) + \frac{\varepsilon (z)\omega_{c
\mathbf{q}}^2}{c^2}f_\mathbf{q}(z)=0
\end{equation}

According to the boundary conditions for the electromagnetic
field, $f_\mathbf{q}$ and $\varepsilon
(z)^{-1}\mathrm{d}f_\mathbf{q}/\mathrm{d}z$ must be continuous at
the interfaces between the different layers, and $f_\mathbf{q}(z)$
must vanish for $z\rightarrow \pm \infty$. Such a function has
been illustrated in Figure \ref{Fig1}(a). Equation
\pref{f_q_equation} will have in general multiple solutions,
describing different modes labelled by a discrete index $j$, that
can be added in the notations if necessary.

One or two boundaries of the multilayer system can be metallic. In
the mid- and far- infrared frequency range, the metals are
described by a very large negative real dielectric constant
$\varepsilon_M<0$, and the electromagnetic field density in the
metallic layers is vanishing. Therefore, in order to have a
consistent quantification scheme for the electromagnetic field,
without the burden of quantifying the conducting electrons in the
metallic boundaries, we neglect the electromagnetic density energy
in the metallic regions in the integrals \pref{H_p0} and
\pref{H_int}. However, the finite dielectric constant
$\varepsilon_M$ can be taken into account through the dispersion
relation of the guided modes, as for the example provided in
Appendix \ref{AppNormEMfield}.

With the use of the above expressions, the Hamiltonian \pref{H_p0}
of the free electromagnetic field takes the standard form:

\begin{equation}\label{H_p}
\hat{H}_p = \sum_{\mathbf{q}}\hbar \omega_{c
\mathbf{q}}(a_\mathbf{q}^\dagger a_\mathbf{q} + 1/2)
\end{equation}

We now apply this formalism to a particular system. For mid-IR and
THz frequencies the simplest confining system is provided by the
$\mathrm{TM}_0$ mode guided between two metallic plates, separated
by a distance $L_{\mathrm{cav}}$, as illustrated in Fig.
\ref{Fig1}(b). The semiconductor with a dielectric constant
$\varepsilon$ fills the space between the plates. In this case, in
the limit of a perfect metallic boundaries we have
$f_\mathbf{q}(z)=1$ and $L_\mathbf{q}=L_{\mathrm{cav}}$ and the
dispersion relation becomes $\mathbf{q}^2=\varepsilon \omega_{c
\mathbf{q}}^2/c^2$ (Fig. \ref{Fig1}(b)). The in-plane
component of the displacement field vanishes,
$\mathbf{\hat{D}}_{\parallel}=\mathbf{0}$ and the remaining
non-zero $z$-component is:

\begin{equation}\label{Dz_TM0}
\mathbf{\hat{D}}_z= i\mathbf{e}_z \sum_\mathbf{q}
\sqrt{\frac{\varepsilon \varepsilon_0 \hbar \omega_{c
\mathbf{q}}}{2SL_{\mathrm{cav}}}}e^{i\mathbf{q}\mathbf{r}_{\parallel}}
(a_\mathbf{q}-a_{-\mathbf{q}}^\dagger)
\end{equation}

In the rest of the paper, we consider exclusively the
$\mathrm{TM}_0$ mode, which simplifies greatly the calculations
without a loss of generality. In section \ref{Sec_0D} we also
consider the case of  double metal zero-dimensional (0D)
microcavities in which we have added a lateral confinement.
In these structures the propagation wavevector $\mathbf{q}$
of the $\mathrm{TM}_0$ becomes quantized.

\subsection{Electronic Hamiltonian}\label{Sec_H_e}

\begin{figure}
\includegraphics[scale=0.16]{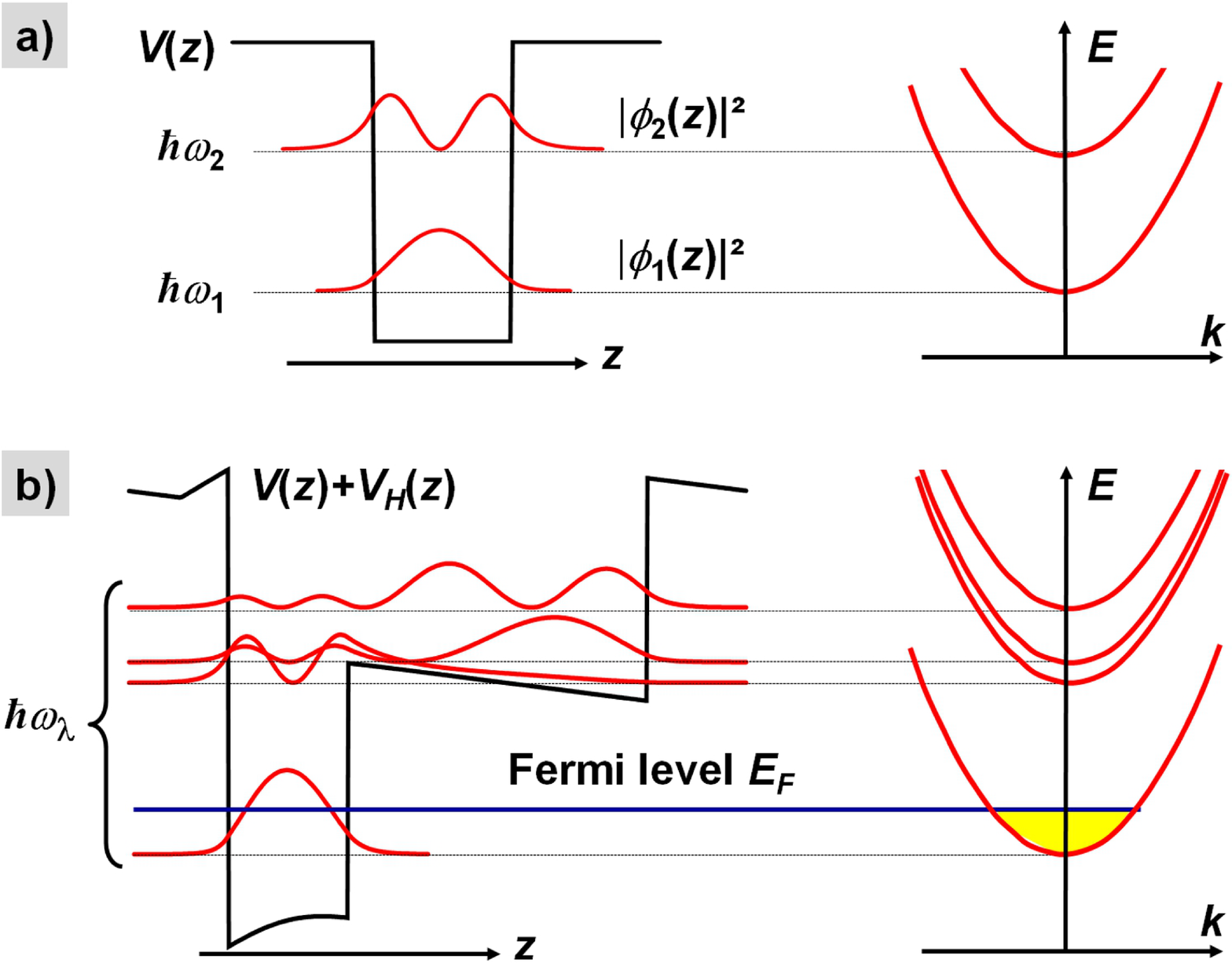}
\caption {(a) Typical quantum heterostructure potential $V(z)$: a
quantum well. There a two bound levels with wavefunctions
$\phi_{1,2}(z)$ (b) General potential of a doped heterostructure,
including the Hartree correction $V_H(z)$ due to the static
Coulomb interactions.} \label{Fig2}
\end{figure}

In a semiconductor heterostructure the band offsets between
different semiconductor layers provide a confining potential
$V(z)$ in the growth axis. Typical example of a confining
heterostructure, the quantum well, is sketched in Figure
\ref{Fig2}(a), and a general heterostructure is
illustrated in Figure \ref{Fig2}(b).
We consider the case where the Fermi level
$E_F$ lies into the conduction band, due to intentional doping,
which leads to the formation of a bi-dimensional electron gas
\cite{Ando_Fowler_Stern_1982} (figure \ref{Fig2}(b)). The
potential $V(z)$ yields discrete energy levels $\hbar
\omega_\lambda$, labelled by an integer index $\lambda$. Electrons
are free to move in the plane perpendicular to the growth axis,
and we denote by $\hbar \mathbf{k}$ their in-plane momentum. This
free movement is described by the parabolic subbands illustrated
in Figures \ref{Fig2}(a) and \ref{Fig2}(b), and the total energy of an
electron in the subband $\lambda$ is:

\begin{equation}
\hbar \omega_{\lambda \mathbf{k}} =\hbar \omega_\lambda +
\frac{\hbar^2 \mathbf{k}^2 }{2m^\ast}
\end{equation}

with $m^\ast$ the effective electron mas.

In a highly doped heterostructure, in order to determine the
confining energies $\hbar \omega_\lambda$, one must take into
account not only the heterostructure potential, but also the
Coulomb interaction between the charges. In the Hartree
approximation, this static Coulomb interactions are described by a
self-consistent potential $V_H(z)$ due to the presence of
electrons and ionised impurities  \cite{Ando_Fowler_Stern_1982,
Book_Helm}. The potential $V_H(z)$ depends on the envelope
wavefunctions $\phi_\lambda(z)$. Therefore the energies $\hbar
\omega_\lambda$ and wavefunction $\phi_\lambda(z)$ are determined
altogether by solving the one-particle Schr\"odinger equation with a
total potential $V(z)+V_H(z)$ self-consistently with a Poisson
problem \cite{Ando_Fowler_Stern_1982}:

\begin{eqnarray}\label{1partSchro}
\Big{[}-\frac{\hbar^2}{2m^\ast}\frac{d^2}{dz^2} +
V(z)+V_H(z)\Big{]}\phi_\lambda(z)= \hbar \omega_\lambda
\phi_\lambda(z)
\\
\frac{d^2 V_H(z)}{dz^2} = -\frac{e^2}{\varepsilon
\varepsilon_0}\Big{[} \rho(z) - N_d(z) \Big{]}
\\
\rho (z) = \frac{m^\ast}{\pi \hbar^2}\sum_\lambda
N_\lambda|\phi_\lambda(z)|^2
\end{eqnarray}

Here $\rho(z)$ the electronic density, $N_d(z)$ is the dopant
density, and $N_\lambda$ is the population of the $\lambda$-th
subband. The exchange-correlation effect has been neglected in the
above equations. This set of equations leads to one-particle
quantum states $|\lambda, \mathbf{k} \rangle$ with normalized
wavefunctions:

\begin{equation}\label{WaveFunc}
\langle \mathbf{r}|\lambda, \mathbf{k}
\rangle=\phi_\lambda(z)\frac{1}{\sqrt{S}}\exp (i
\mathbf{k}\mathbf{r}_{\parallel})
\end{equation}

The corresponding fermionic destruction and creation operators are
$c_{\lambda \mathbf{k}}$ and $c_{\lambda \mathbf{k}}^\dagger$. The
electronic Hamiltonian acquires the one-particle form:

\begin{equation}\label{H_e}
\hat{H}_e = \sum_{\lambda \mathbf{k}}\hbar \omega_{\lambda
\mathbf{k}}c_{\lambda \mathbf{k}}^\dagger c_{\lambda \mathbf{k}}
\end{equation}

At this point we have determined the stationary state of the
system. The effects of the Coulomb interaction that we have
considered so far are the static effects arising from the
inhomogeneous spatial distribution of charges. These static
effects have been lumped into the one-particle subband energies
$\hbar \omega_{\lambda \mathbf{k}}$, through the self-consistent
set of equation (13)-(15).

In the next section we use the stationary basis of single particle
states $\phi_\lambda(z)$ in order to define the microscopic
polarization density of the electron gas. This will allow us to
study the excited collective states, coupled with light, and to
recover the dynamic effects of the Coulomb interaction, such as
the depolarization shift.

\subsection{Microscopic expression of the Polarization}\label{Sec_MicroDefP}

Usually, for the studies of the interaction of intersubband
transitions with light, one does not define a polarization
operator as an independent local quantity
$\mathbf{\hat{P}}(\mathbf{r})$, but rather computes the linear
response of the electronic system due to the solicitation of an
external harmonic electric field. The linear response is described by the
frequency dependent non-local susceptibility
$\chi(\omega)$\cite{Dahl_Sham_1977} which is computed from the
current-current correlation function through the Kubo formula
\cite{Wendler_Kandler_1993}. The collective excitations of the
system are then obtained from the isolated poles of
$\chi(\omega)$.

In order to use the electrical dipole gauge, as formulated by the
interaction Hamiltonian \pref{H_int}, we need to express the local
polarization operator $\mathbf{\hat{P}}(\mathbf{r})$ as an
independent quantity. Classically, the local polarization
$\mathbf{P}(\mathbf{r})$ is defined as the average dipole
moment of the charge distribution over some microscopic volume
\cite{Book_Landau_ECM}. For instance, in the case of the square
quantum well with a thickness $L_{\mathrm{QW}}$ illustrated in
Figure \ref{Fig2}(a), such volume could be the volume of the
quantum well. However, it is difficult to apply such a definition
for an arbitrary heterostructure, such as the one depicted in
Figure \ref{Fig2}(b), where the spatial extension of the
confinement potential varies with the energy. In this case, the
averaging volume becomes an arbitrary quantity. Since the
microscopic intersubband dipole is defined from the electronic
wavefunctions $\phi_\lambda(z)$, the truly microscopic expression
of $\mathbf{\hat{P}}(\mathbf{r})$ should involve directly
$\phi_\lambda(z)$.

Similar problem has been encountered in the attempts to define the
static polarization in ferromagnetic materials as function of the
electronic wavefunctions and nuclei distributions
\cite{Resta_2010}. Then the idea is to define the polarization as
the time integral of the microscopic current arising during the
adiabatic switch from one configuration of electrons and nuclei to
another.

In our case, the microscopic current which corresponds to the
intersubband transitions is a rapidly oscillating function at the
frequencies of the transitions. Then we define the polarization in
such a way that its time evolution under the full Hamiltonian
would lead to a microscopic current:

\begin{equation}\label{def_Pol}
\frac{d\mathbf{\hat{P}}(\mathbf{r})}{dt} = \frac{1}{i\hbar}[
\mathbf{\hat{P}}(\mathbf{r}),\hat{H}] =
\mathbf{\hat{j}}(\mathbf{r})
\end{equation}

This definition is valid in the absence of magnetic interactions.
Since the polarization operator commutes with the electrical
displacement field the evolution of $\mathbf{P}(\mathbf{r})$ is
driven only by the electronic part of Hamiltonian \pref{H_e}.
The expression of the total current operator in the PZW gauge is:

\begin{equation}
\mathbf{\hat{j}}(\mathbf{r}) = \frac{i\hbar e}{2
m^\ast}[\hat{\Psi}^\dagger(\mathbf{r})\nabla_\mathbf{r}\hat{\Psi}(\mathbf{r})-
\nabla_\mathbf{r}\hat{\Psi}^\dagger(\mathbf{r})\hat{\Psi}(\mathbf{r})]
\end{equation}

Note that the paramagnetic term, proportional to the vector
potential $\mathbf{A}$, is absent, since in the PZW gauge the
momentum of the particles is expressed as a function of their
velocity only \cite{Book_Cohen_Ph_At}. Here we have introduced the
field operator $\hat{\Psi}(\mathbf{r})$, constructed from the
one-particle wavefunctions \pref{WaveFunc}:

\begin{equation}\label{Field_op}
\hat{\Psi}(z, \mathbf{r}_{\parallel}) = \sum_{\lambda
\mathbf{k}}c_{\lambda \mathbf{k}}\phi_\lambda(z)\exp (i
\mathbf{k}\mathbf{r}_{\parallel})/\sqrt{S}
\end{equation}

Since we are interested in the intersubband transitions only, we
shall consider solely the $z$-component of the current. The latter
is readily expressed as:

\begin{eqnarray}
\hat{j}_z(\mathbf{r}) = \frac{i\hbar e}{2S m^\ast}\sum_{\lambda >
\mu, \mathbf{q}}\xi_{\lambda \mu}(z)
e^{i\mathbf{q}\mathbf{r}_{\parallel}} [B_{\lambda \mu
\mathbf{q}}-B^\dagger_{\lambda \mu -\mathbf{q}}]
\end{eqnarray}

with the following definitions:

\begin{equation}\label{B-op}
B^\dagger_{\lambda \mu \mathbf{q}} =
\sum_{\mathbf{k}}c^\dagger_{\lambda \mathbf{k}+\mathbf{q}}c_{\mu
\mathbf{k}}
\end{equation}

\begin{equation}\label{xi-density}
\xi_{\lambda \mu} (z) =\phi_\lambda(z)\partial_z
\phi_\mu(z)-\phi_\mu(z)\partial_z \phi_\lambda(z)
\end{equation}

For simplicity, we have chosen the envelope wavefunctions of the
bound states $\phi_\lambda(z)$ to be real. Note that the
intrasubband contribution all vanish since $\xi_{\lambda \lambda}
(z)=0$ according to the above definition.

In order to establish the polarization operator in the long
wavelength limit, we first compute the commutators of the
B-operators \pref{B-op} with the Hamiltonian \pref{H_e}:

\begin{equation}\label{Com_0}
[B^\dagger_{\lambda \mu \mathbf{q}}, \hat{H}_e] =
-\sum_{\mathbf{k}}\hbar (\omega_{\lambda
\mathbf{k}+\mathbf{q}}-\omega_{\mu \mathbf{k}})c^\dagger_{\lambda
\mathbf{k}+\mathbf{q}}c_{\mu \mathbf{k}}
\end{equation}

In the long wavelength limit the excitation wavevector $\mathbf{q}$
is small compared to the typical electron wavevectors $\mathbf{k}$.
Since we assumed parabolic bands, we can write:

\begin{equation}
\omega_{\lambda \mathbf{k}+\mathbf{q}}-\omega_{\mu \mathbf{k}}
\approx \omega_\lambda - \omega_\mu \equiv \omega_{\lambda \mu}
\end{equation}

Then \pref{Com_0} becomes:

\begin{equation}\label{Com_B}
[B^\dagger_{\lambda \mu \mathbf{q}},\hat{H}_e] = -\hbar
\omega_{\lambda \mu} B^\dagger_{\lambda \mu \mathbf{q}}
\end{equation}

This commutation relation indicates that the polarization density
operator satisfying \pref{def_Pol} is:

\begin{eqnarray}\label{P_density}
\hat{P}_z(\mathbf{r}) = \frac{\hbar e}{2S m^\ast} \sum_{\lambda >
\mu, \mathbf{q}}\frac{\xi_{\lambda \mu} (z)}{\omega_{\lambda
\mu}}e^{i\mathbf{q}\mathbf{r}_{\parallel}}[B^\dagger_{\lambda \mu
\mathbf{-q}} +B_{\lambda \mu \mathbf{q}}]
\end{eqnarray}

This is the expression to be used in the interaction Hamiltonian
$\hat{H}_{\mathrm{int}}$ \pref{H_int}. It is clear that this
expression satisfies the requirement stated in the beginning of
this section. Indeed, it is a local function of space through the
the microscopic current density $\xi_{\lambda \mu} (z)$. The size
of the confining potential enters only implicitly, through the
wavefunctions that construct $\xi_{\lambda \mu} (z)$ and we do not
need to define any arbitrary averaging volume.

To express the interaction Hamiltonian $\hat{H}_{\mathrm{int}}$ as
a function of the electronic polarization, it is convenient to
split it into two parts:

\begin{equation}
\hat{H}_{\mathrm{int}}=\hat{H}_{\mathrm{I1}}+\hat{H}_{\mathrm{I2}}
\end{equation}

with $\hat{H}_{\mathrm{I1}}$ the linear part and
$\hat{H}_{\mathrm{I2}}$ the part quadratic in the polarization.
The expression of $\hat{H}_{\mathrm{I1}}$ in the long wavelength
limit, for the case of the $\mathrm{TM}_0$ mode, is readily
obtained with the help of \pref{Dz_TM0}:

\begin{eqnarray}\label{H_I1}
\hat{H}_{\mathrm{I1}} = i\sum_{\lambda > \mu, \mathbf{q}}
\sqrt{\frac{ \hbar \omega_{c \mathbf{q}}}{2\varepsilon
\varepsilon_0 S
L_{\mathrm{cav}}}}ez_{\lambda \mu} \times \nonumber\\
(a^\dagger_\mathbf{q}-a_{\mathbf{-q}})(B^\dagger_{\lambda \mu
-\mathbf{q}}+B_{\lambda \mu \mathbf{q}})
\end{eqnarray}

Here we have introduced the dipole matrix element $z_{\lambda \mu}
= \langle \phi_\lambda|z|\phi_\mu \rangle $ of the transition $\mu
\rightarrow \lambda$, and made use of the following identity:

\begin{equation}\label{current_identity}
\int_{-\infty}^{+\infty} \xi_{\lambda \mu} (z) dz =\frac{2 m^\ast
\omega_{\lambda \mu }}{\hbar} z_{\lambda \mu}
\end{equation}

This identity guarantees the equivalence between the current and
dipole matrix elements, which are more commonly used in studies of
intersubband transitions \cite{Book_Helm, Sirtori_1994}.

The quadratic term of the interaction Hamiltonian becomes:

\begin{eqnarray}\label{H_I2}
\hat{H}_{\mathrm{I2}} = \frac{e^2 \hbar^2}{8 \varepsilon
\varepsilon_0 S {m^\ast}^2}  \sum_{\lambda > \mu, \lambda ' > \mu
', \mathbf{q}}\frac{I_{\lambda \mu, \lambda '  \mu
'}}{\omega_{\lambda \mu} \omega_{\lambda' \mu'}}\times \nonumber\\
(B^\dagger_{\lambda \mu \mathbf{q}}+B_{\lambda \mu
-\mathbf{q}})(B^\dagger_{\lambda '\mu ' -\mathbf{q}}+B_{\lambda '
\mu ' \mathbf{q}})
\end{eqnarray}

where $I_{\lambda \mu, \lambda '  \mu '}$ denotes the
current-current overlap integral:

\begin{equation}\label{Current_Corr}
I_{\lambda \mu, \lambda '  \mu '}= \int_{-\infty}^{+\infty}
\xi_{\lambda \mu} (z) \xi_{\lambda ' \mu '} (z) dz
\end{equation}

Note that the expressions derived above are exact and general,
except for the long-wavelength approximation that is satisfied for
the majority of experiments with intersubband devices. For
instance, they will apply for a heterostructure featuring
population inversion. In the following, we are interested in the
specific case of non-inverted (thermalized) subbands.

\section{Bosonized Plasma Hamiltonian}\label{Part_II}

\subsection{Bright and Dark states}\label{Bright_and_Dark}

In order to further study  the interaction between the light and the
thermalized subbands, we replace the fermionic Hamiltonian
$\hat{H}_e$ by an effective bosonic Hamiltonian which contains
only the polarization degrees of freedom. In this way, we obtain a
fully diagonalizable Hopfield-like model. To carry out this
approach we need to replace the B-operators \pref{B-op} by
effective bosonic operators. For this purpose we compute the
commutator:

\begin{eqnarray}
[B_{\lambda \mu \mathbf{q}}, B^\dagger_{\lambda \mu \mathbf{q}}] =
\nonumber\\ \sum_{\mathbf{k}}(c^\dagger_{\mu \mathbf{k}}c_{\mu
\mathbf{k}}-c^\dagger_{\lambda \mathbf{k+q}}c_{\lambda
\mathbf{k+q}}) = \hat{N}_\mu - \hat{N}_\lambda
\end{eqnarray}

We recognize the difference between the number operators
$\hat{N}_\mu$ and $\hat{N}_\lambda$ of the respective subbands.
Let us define normalized operators through the relation:

\begin{equation}\label{def_b}
B^\dagger_{\lambda \mu \mathbf{q}} = \sqrt{\Delta N_{\lambda
\mu}}b^\dagger_{\lambda \mu \mathbf{q}}
\end{equation}

with:

\begin{equation}
\Delta N_{\lambda \mu} = \langle\hat{N}_\mu\rangle -
\langle\hat{N}_\lambda\rangle
\end{equation}

Here the mean values of the operators are computed in the, say,
thermal state arising from the Fermi-Dirac statistics. In the
limit of weakly excited system, the operators $b^\dagger_{\lambda
\mu \mathbf{q}}$ defined above obey the bosonic commutation rules
\cite{DeLiberato_Ciuti_PRL2009}:

\begin{equation}\label{Com_b}
[b_{\lambda \mu \mathbf{q}}, b^\dagger_{\lambda \mu \mathbf{q'}}]
=\delta_{\mathbf{q}, \mathbf{q'}}
\end{equation}

Taking into account \pref{Com_b}, \pref{def_b} and \pref{Com_B} we
replace the fermionic Hamiltonian $\hat{H}_e$ \pref{H_e} by an
effective bosonic Hamiltonian which yields exactly the same time
evolution of the weakly excited system:

\begin{equation}\label{H'_e}
\hat{H}'_e = \sum_{\lambda > \mu, \mathbf{q}}\hbar \omega_{\lambda
\mu} b^\dagger_{\lambda \mu \mathbf{q}}b_{\lambda \mu \mathbf{q}}
\end{equation}

\begin{figure}
\includegraphics[scale=0.15]{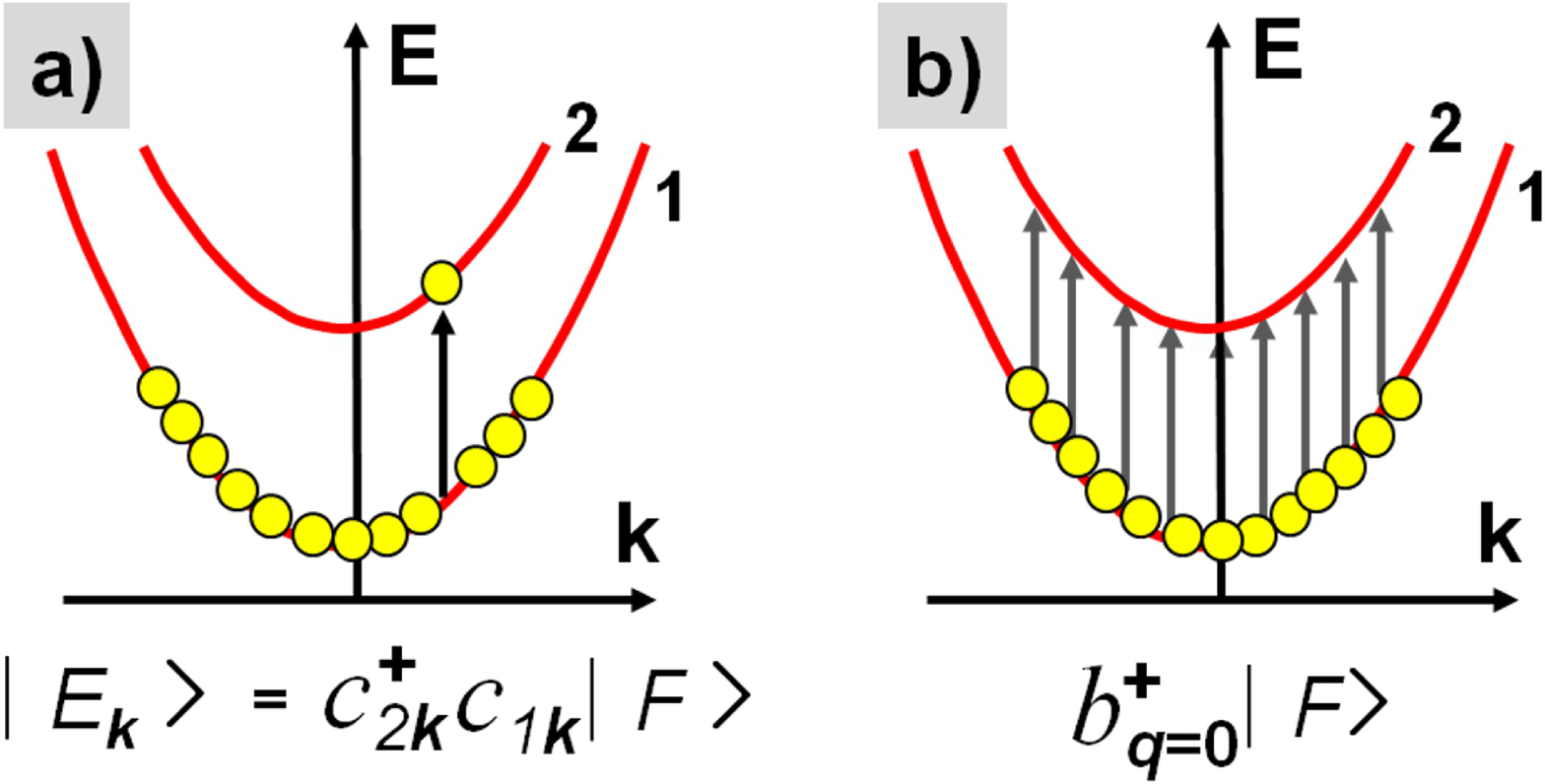}
\caption {a) One particle excitation between two subbands. The
fundamental subband is filled with $N$ electrons at $T=0$K. b)
Bright excitation.} \label{Fig3}
\end{figure}

The bosonic operators introduced here describe the only
intersubband excitations that couple with light
\cite{Ciuti_PhysRevB_2005}, and they are therefore called "bright"
states. To illustrate the bright states, we consider the system to
be in the electric quantum limit at $T=0$ K, where only the first
subband ($\lambda =1$) is occupied by $N$ electrons. The
fundamental state is then given by $|F\rangle =
\Pi_{|\mathbf{k}|<k_F} c^\dagger_{1\mathbf{k}}|0\rangle$, where
$k_F$ is the Fermi wavevector. The lowest energy single particle
excitations can be spanned on the basis of states:

\begin{equation}
|E_\mathbf{k}\rangle =
c^\dagger_{2\mathbf{k}}c_{1\mathbf{k}}|F\rangle
\end{equation}

These basis states are eigenstates of the electronic Hamiltonian
$\hat{H}_e$ with an eigenenergies $\hbar
\omega_{21}+E_\mathrm{fond}$, where $E_\mathrm{fond}$ is the total
energy of the ground state. In the following, the energy scale is
reset so that $E_\mathrm{fond}=0$. The excitations
$|E_\mathbf{k}\rangle$ are illustrated in Figure
\ref{Fig3}(a). For simplicity, in our example we consider
only vertical transitions $\mathbf{q}=\mathbf{0}$. The dipole
moment operator $\hat{d}$ between the subbands 1 and 2 is:

\begin{equation}
\hat{d} = z_{12}
\sum_\mathbf{k}(c^\dagger_{2\mathbf{k}}c_{1\mathbf{k}}+c^\dagger_{1\mathbf{k}}c_{2\mathbf{k}})
\end{equation}

It is easy to show that each excitation $|E_\mathbf{k}\rangle$
holds a dipole $z_{12}=\langle F|\hat{d}| E_\mathbf{k} \rangle$.
We can, however, change the basis by using any appropriate linear
combinations of $|E_\mathbf{k}\rangle$'s:

\begin{equation}
|B_a \rangle = \sum_{\mathbf{k}} \beta_{a,\mathbf{k}}|
E_\mathbf{k} \rangle
\end{equation}

\begin{equation}\label{norm_beta}
\sum_{\mathbf{k}} \beta_{a,\mathbf{k}}\beta^\ast_{b,\mathbf{k}} =
\delta_{a,b}
\end{equation}

\begin{equation}\label{dipole_beta}
\langle F|\hat{d}| B_a \rangle = z_{12} \sum_\mathbf{k}
\beta_{a,\mathbf{k}}
\end{equation}

Here $a,b=1..N$ label the new basis states. The coefficients
$\beta_{a,\mathbf{k}}$ are arbitrary within the normalization
condition \pref{norm_beta}, but let us define the $a=1$ state so
that all $\beta_{1,\mathbf{k}}$ are equal. Because of
\pref{norm_beta} we then have $\beta_{1,\mathbf{k}}=1/\sqrt{N}$
and therefore $|B_1 \rangle$ is exactly the state created by the
bosonic operator introduced above: $|B_1 \rangle
=b^\dagger_{\mathbf{q}=0}|F\rangle$. This state, which is a
coherent superposition of all possible single particle states with
equal amplitudes $1/\sqrt{N}$ is illustrated in Figure
\ref{Fig3} (b). Next, because of \pref{dipole_beta} and
\pref{norm_beta} we have:

\begin{eqnarray}
\langle F|\hat{d}| B_1 \rangle = z_{12}\sqrt{N}, \phantom{QQ}
\langle F|\hat{d}| B_{a \neq 1} \rangle =0
\end{eqnarray}

These formulas express the fact that the whole oscillator strength
of the system is hold by the single state
$b^\dagger_{\mathbf{q}=0}|F\rangle$, whereas all other $N-1$
states, that are orthogonal to it, have a zero oscillator strength
and do not couple with light.

At a first glance it might seem that the emergence of the
superradiant state $b^\dagger_{\mathbf{q}=0}|F\rangle$ is due to a
mere change of the basis, and that this state does not have any
particular physical meaning. Indeed, this state is perfectly
degenerate with the dark states as respect to the electronic
Hamiltonian $\hat{H}_e$. However, the inclusion of the interaction
Hamiltonian \pref{H_int} lifts the degeneracy between the bright
superradiant state and the dark states $|B_{a \neq 1} \rangle$,
since both the quadratic $\hat{H}_{I2}$ and the linear
$\hat{H}_{I1}$ parts renormalize the energy of the bright state.
The quadratic part $\hat{H}_{I2}$ leads to a blue shift, as
explained in the next section, and the linear part to the
emergence of two polariton states. Neither of $\hat{H}_{I1}$ or
$\hat{H}_{I2}$ acts on the dark states, which therefore remain at
the energy of the bare intersubband transition $\hbar
\omega_{21}$. These heavily degenerated dark states hinder the
efficiency of the electronic injection into the polariton states
\cite{DeLiberato_Ciuti_Tunnel_2009}.

\subsection{Plasma Hamiltonian}\label{Sec_PlasmaH}

The procedure of bosonization described in the previous section
assigns bosonic operators $b_{\lambda \mu \mathbf{q}}$ and
$b^\dagger_{\lambda \mu \mathbf{q}}$ to each intersubband
transition $\mu \rightarrow \lambda$. Each transition now enters the
bosonized Hamiltonian with an energy $\hbar \omega_{\lambda \mu}$.
Therefore, in order to avoid cumbersome notations, from now on we
label each transition $\mu \rightarrow \lambda$ with a single Greek index
$\alpha$, i.e. $\alpha \equiv [\lambda ,\mu]$. This means that now
we count the number of excitations in the system, instead of the
number of subband states.

We now seek to express the interaction Hamiltonian as a function
of the bosonic operators $b_{\alpha \mathbf{q}}$ and
$b^\dagger_{\alpha \mathbf{q}}$. To this end, we introduce the
plasma frequencies $\omega_{P\alpha}$ through the formula:

\begin{equation}\label{Plasma_Freq}
\omega^2_{P\alpha} = \frac{e^2 \Delta N_\alpha}{\varepsilon
\varepsilon_0 m^\ast S L^{\alpha}_{\mathrm{eff}}}
\end{equation}

Here $L^{\alpha}_{\mathrm{eff}}$ is the effective length
introduced by Vinter and Tsui \cite{Ando_Fowler_Stern_1982}. This
length is a function on the current-current correlation function
introduced by equation \pref{Current_Corr}, and describes the
spatial extension of the microscopic current density of the
intersubband transition $\alpha$:

\begin{equation}\label{Leff}
L^{\alpha}_{\mathrm{eff}} = \frac{2m^\ast
\omega_{\alpha}}{\hbar}\frac{1}{I_{\alpha, \alpha}}
\end{equation}

We will also make use of the transition oscillator strength:

\begin{equation}\label{ostrength}
f^o_\alpha=\frac{2m^\ast \omega_\alpha}{\hbar}z_\alpha^2
\end{equation}

The linear part of the interaction Hamiltonian then becomes:

\begin{eqnarray}
\hat{H}_{\mathrm{I1}} =  i\sum_{\alpha, \mathbf{q}} \frac{\hbar
\omega_{P\alpha}}{2}\sqrt{\frac{\omega_{c
\mathbf{q}}}{\omega_\alpha}f^o_{\alpha}f^w_{\alpha}}\times
\nonumber\\(a^\dagger_\mathbf{q}-a_{-\mathbf{q}})(b^\dagger_{\alpha
-\mathbf{q}}+b_{\alpha \mathbf{q}})
\end{eqnarray}

Here $f^w_{\alpha}$ is the overlap factor between the cavity mode
and the current distribution of the transition $\alpha$, defined
as:

\begin{equation}\label{def_fw_TM0}
f^w_{\alpha} = L^{\alpha}_{\mathrm{eff}}/L_{\mathrm{cav}}
\end{equation}

This definition, introduced here for simplicity for the special case
of a $\mathrm{TM}_0$ mode can be generalized for a mode with
an arbitrary shape, as shown in the end of this section.

We now turn to the quadratic part $\hat{H}_{I2}$ of the
interaction Hamiltonian. It is expressed as a sum over pairs of
transitions $\lambda > \mu, \lambda' > \mu' = \alpha, \beta$. Let
us consider first the terms that correspond to the same
transition, $\alpha=\beta$. Combining these terms with the
electronic Hamiltonian \pref{H'_e} we have:

\begin{eqnarray}\label{Hselfenergy}
\hat{H}'_e + \hat{H}_{I2}(\alpha = \beta) = \sum_{\alpha,
\mathbf{q}} [\hbar \omega_\alpha b^\dagger_{\alpha
\mathbf{q}}b_{\alpha \mathbf{q}}\nonumber\\+ \frac{\hbar
\omega_{P\alpha}^2}{4\omega_\alpha} (b^\dagger_{\alpha
\mathbf{q}}+b_{\alpha -\mathbf{q}})(b^\dagger_{\alpha
-\mathbf{q}}+b_{\alpha \mathbf{q}})]
\end{eqnarray}

This quadratic Hamiltonian can be diagonalized with the Bogoliubov
transformation \cite{Book_Schawbl}, by introducing new bosonic
operators $p_{\alpha \mathbf{q}}$ which satisfy:

\begin{equation}
[p_{\alpha \mathbf{q}},\hat{H}'_e + \hat{H}_{I2}(\alpha =
\beta)]=\hbar \widetilde{\omega}_{\alpha} p_{\alpha \mathbf{q}}
\end{equation}

where $\widetilde{\omega}_\alpha$ denotes the new eigenvalues.
This diagonalization procedure yields the following results:

\begin{equation}\label{depol_shift}
\widetilde{\omega}_\alpha = \sqrt{\omega_\alpha^2+\omega_{P
\alpha}^2}
\end{equation}

\begin{equation}\label{p_ops}
p_{\alpha \mathbf{q}} =
\frac{\widetilde{\omega}_{\alpha}+\omega_{\alpha}}{2\sqrt{\widetilde{\omega}_{\alpha}\omega_{\alpha}}}b_{\alpha
\mathbf{q}}
+\frac{\widetilde{\omega}_{\alpha}-\omega_{\alpha}}{2\sqrt{\widetilde{\omega}_{\alpha}\omega_{\alpha}}}b^\dagger_{\alpha
\mathbf{-q}}
\end{equation}

In equation \pref{depol_shift} the new eigenvalue
$\widetilde{\omega}_\alpha$ is exactly the frequency of the
collective mode of the bi-dimensional electron gas known as the
"intersubband plasmon" \cite{Wendler_Kandler_1993}. Using
\pref{Plasma_Freq} and \pref{p_ops} to express the remaining
$\alpha \neq \beta$ terms of $\hat{H}_{I2}$ we arrive at the full
Hamiltonian, which is now expressed in terms of the collective
plasmonic operators:

\begin{eqnarray}\label{Hplasma_q}
\hat{H} =  \sum_{\alpha, \mathbf{q}}\hbar
\widetilde{\omega}_\alpha p^\dagger_{\alpha \mathbf{q}}p_{\alpha
\mathbf{q}} +\sum_{\mathbf{q}}\hbar
\omega_{c \mathbf{q}}(a_\mathbf{q}^\dagger a_\mathbf{q} + 1/2)\nonumber\\
+i\sum_{\alpha, \mathbf{q}} \hbar\Omega_{\alpha \mathbf{q}}
(a^\dagger_\mathbf{q}-a_{-\mathbf{q}})(p^\dagger_{\alpha
-\mathbf{q}}+p_{\alpha \mathbf{q}})\nonumber\\
+\sum_{\alpha \neq \beta, \mathbf{q}} \hbar \Xi_{\alpha \beta}
(p^\dagger_{\alpha \mathbf{q}}+p_{\alpha
-\mathbf{q}})(p^\dagger_{\beta -\mathbf{q}}+p_{\beta \mathbf{q}})
\end{eqnarray}

Here we have introduced the light-matter coupling constant in the
dipole gauge:

\begin{equation} \label{Coupling_Const_Dipgauge}
\Omega_{\alpha \mathbf{q}} =
\frac{\omega_{P\alpha}}{2}\sqrt{\frac{\omega_{c
\mathbf{q}}}{\widetilde{\omega}_{\alpha}}f_{\alpha}^o
f_{\alpha}^w}
\end{equation}

The quantity $\Xi_{\alpha \beta}$ is the plasmon-plasmon coupling
constant:

\begin{equation}\label{Const-Plasmon-Plasmon}
\Xi_{\alpha \beta} = \frac{\omega_{P\alpha} \omega_{P\beta}}{4
\sqrt{\widetilde{\omega}_{\alpha} \widetilde{\omega}_{\beta}}}
C_{\alpha \beta}
\end{equation}

where the coefficient $C_{\alpha \beta}$ is the plasmon-plasmon
correlation coefficient, defined as:

\begin{equation}\label{Plasmon-Plasmon}
C_{\alpha \beta} = \frac{I_{\alpha,\beta}}{\sqrt{I_{\alpha,
\alpha}I_{\beta, \beta}}} = \frac{\int \xi_{\alpha} (z)
\xi_{\beta} (z) dz}{\sqrt{I_{\alpha, \alpha}I_{\beta, \beta}}}
\end{equation}

The plasma Hamiltonian described in equation \pref{Hplasma_q} is
the central result of this paper. It provides a fully quantum
description of the coupling between the light and the coherent
collective intersubband modes of a bi-dimensional electron system.
The Hamiltonian \pref{Hplasma_q} contains both the interaction
with the electromagnetic field and the coupling between plasmons
from different subbands. The inter-plasmon coupling is contained
in the coefficients $C_{\alpha \beta}$. With the definition
\pref{Current_Corr} it appears simply as the normalized spatial
overlap between the intersubband currents associated to the
transitions $\alpha$ and $\beta$. This overlap vanishes when the
subbands belong to spatially different quantum wells. On the
contrary, $C_{\alpha \beta}$ takes values close to unity when the
subbands originate from the same quantum well. The coefficients
$C_{\alpha \beta}$ provide thus a convenient description for a
number of cases, from spatially decoupled quantum wells
(tight-binding approximation), to strongly coupled
heterostructures such as a superlattice. When several intersubband
plasmons are present in the system, the Bogoliubov procedure can
be further applied to the inter-plasmon coupling terms of equation
\pref{Hplasma_q} in order to obtain the new normal modes and their
coupling with the light field. These results will be presented in
a separate paper.

We conclude this section by providing a very general expression of
the light-matter coupling constant for an arbitrary shaped guided
mode $f_\mathbf{q}(z)$. To establish this expression we use the
general form the $z$-component of the displacement field
\pref{D_z_general}, instead of the special case of the
$\mathrm{TM}_0$ mode \pref{Dz_TM0}. We then express the linear
coupling term $\hat{H}_{I1}$ by following the same procedure as
described in section \ref{Sec_MicroDefP} and the beginning of this
section. This leads to the result:

\begin{equation} \label{Coupling_Const_General}
\Omega_{\alpha \mathbf{q}} =
\frac{\omega_{P\alpha}}{2\sqrt{\widetilde{\omega}_{\alpha}}}\sqrt{\omega_{c
\mathbf{q}}}\cos \theta_\mathbf{q}C_{\alpha, \mathbf{q}}
\end{equation}

Here we have introduced the normalized current-light overlap
coefficient:

\begin{equation}\label{Plasmon-Light}
C_{\alpha, \mathbf{q}} =\frac{\int
f_\mathbf{q}(z)\xi_{\alpha}(z)dz}{\sqrt{L_\mathbf{q}I_{\alpha,
\alpha}}}
\end{equation}

and the angle $\theta_\mathbf{q}$ is the propagation angle between
the guided mode and the in-plane direction:

\begin{equation}
\cos \theta_\mathbf{q} =
\frac{|\mathbf{q}|c}{\sqrt{\varepsilon}\omega_{c \mathbf{q}}}
\end{equation}

\begin{figure}
\includegraphics[scale=0.125]{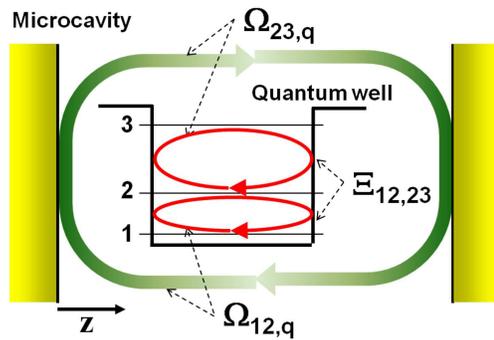}
\caption {Scheme of all possible interactions between a
microcavity and a three level quantum well. The thick arrow
corresponds to the electromagnetic mode and the thin arrows to the
intersubband plasmon modes.} \label{Fig4}
\end{figure}

In the previous expression $\varepsilon = \varepsilon (z \approx
z_\alpha)$ is the background dielectric constant of the media
surrounding the current density $\xi_{\alpha}(z)$. The expressions
\pref{Const-Plasmon-Plasmon} and \pref{Coupling_Const_General}
reveal the striking resemblance between the plasmon-plasmon and
the plasmon-light coupling constants, $\Xi_{\alpha,\beta}$ and
$\Omega_{\alpha,\mathbf{q}}$. Indeed, the two coupling constants
are proportional to a normalized overlap factor, respectively
$C_{\alpha, \beta}$ and $C_{\alpha,\mathbf{q}}$ (equations
\pref{Plasmon-Plasmon} and \pref{Plasmon-Light}). In the same way
as $C_{\alpha, \beta}$ represents the overlap integral between the
two currents arising from transition $\alpha$ and $\beta$,
$C_{\alpha,\mathbf{q}}$ is the overlap between the current
distribution $\alpha$ and the electrical field profile of the
optical mode $f_\mathbf{q}(z)$. Moreover, each plasmon enters the
interaction with a weight factor
$\omega_{P\alpha}/2\sqrt{\widetilde{\omega}_{\alpha}}$. The weight
of the light mode is $\sqrt{\omega_{c \mathbf{q}}}\cos
\theta_\mathbf{q}$, the cosine term expressing the selection rule
for intersubband transitions \cite{Book_Helm}. Therefore, the
interaction between the different plasmons, and the interaction
with the light mode have the same form in the dipole gauge.
Indeed, the Hamiltonian \pref{Hplasma_q} describes a set of
coupled oscillators, one of which is the electromagnetic
resonator, the other being the collective plasmon modes. This is
schematized in in Figure \ref{Fig4}.

In most experimental situations the function $f_\mathbf{q}(z)$
varies slowly at the scale of the intersubband current density
$\xi_\alpha(z)$, and therefore can be taken out from the integral
in equation \pref{Plasmon-Light}. In this case we recover the usual expression 
for the oscillator strength $f^o_{\alpha}$ \cite{Sirtori_1994} using \pref{ostrength} and \pref{current_identity}. This leads to the expression \pref{Coupling_Const_Dipgauge}, where instead of
the definition \pref{def_fw_TM0} we use a generalized filling factor of the form:

\begin{equation}\label{General_fw}
f^w_{\alpha} =
\frac{L^{\alpha}_{\mathrm{eff}}f_{\mathbf{q}}(z\approx
z_\alpha)^2}{L_{\mathbf{q}}}\cos^2 \theta_\mathbf{q}
\end{equation}

(The definition \pref{def_fw_TM0} for the $\mathrm{TM}_0$ mode is
recovered by setting  $\theta_\mathbf{q}=0$ and
$L_{\mathbf{q}}=L_{\mathrm{cav}}$). The dipole gauge provides
therefore a very compact description of the interaction between
the light and intersubband excitations. In particular, the simple
structure of the coupling constants allows to disclose role played
by the collective effect, described by the plasma frequencies
$\omega_{P\alpha}$ and the microcavity geometry, which sets the
overlap coefficients. Finally, in this formulation, the weak
coupling regime is naturally recovered for the case of vanishing
plasmon-light overlap ($C_{\alpha, \mathbf{q}} \rightarrow 0$ or
$f^w_{\alpha} \rightarrow 0$), as expected. In this situation,
which is common for absorption experiments, one measures solely
the collective effects in the intersubband system contained in the
matter part of the quantum Hamiltonian \pref{Hplasma_q}
\cite{Ando_Fowler_Stern_1982,Pinczuk_1993}.

\subsection{Case of 0D microcavities}\label{Sec_0D}

So far we used the expansion of the electromagnetic into the basis
of guided modes labelled by the wavevector $\mathbf{q}$. Our choice
was motivated by the microscopic definition of the polarization
density in section \ref{Sec_MicroDefP}, which is naturally
expanded into elecronic plane waves propagating along the
heterostucture slab (see, for instance, \pref{B-op}).

The approach developed here allows the rigorous quantum
description of 0D microcavities that confine the electromagnetic
field into all tree dimensions of space. Such systems have been
recently employed for the study of light-matter coupling with
intersubband transitions \cite{Todorov_PRL2010, geiser_2010}. In
this case the possible polarization excitations will be determined
by the quantizing conditions for $\mathbf{q}$ imposed by the
microcavity boundaries.

We start by expanding the electrical displacement into laterally
localized $\mathrm{TM}_0$-like modes:

\begin{eqnarray}\label{D_m}
\mathbf{\hat{D}}(\mathbf{r}) =
\mathbf{\hat{D}}(z,\mathbf{r}_\parallel)=
\nonumber\\
i\mathbf{e}_z \sum_m \sqrt{\frac{\varepsilon \varepsilon_0 \hbar
\omega_{c m}}{2SL_{\mathrm{cav}}}}u_m(\mathbf{r}_{\parallel})
(a_m^\dagger - a_m)
\end{eqnarray}

The index $m$ labels the discrete cavity modes and the set of
functions $u_m(\mathbf{r}_{\parallel})$ describe the lateral shape
of the modes. They are normalized such as:

\begin{equation}
\iint u_m(\mathbf{r}_{\parallel})u_{m'}(\mathbf{r}_{\parallel})d^2
\mathbf{r}_{\parallel} = S \delta_{m m'}
\end{equation}

We must now provide the expansion the polarization density
\pref{P_density} into the new basis of localized electromagnetic
modes. To this end, it is very convenient to use the quantum
mechanical notations $|\mathbf{q}\rangle$ and $|u_m\rangle$ so
that $\langle \mathbf{r}_{\parallel}|\mathbf{q}\rangle =
\exp(i\mathrm{q}\mathbf{r}_{\parallel})/\sqrt{S}$ and $\langle
\mathbf{r}_{\parallel}|u_m\rangle =u_m(\mathbf{r}_{\parallel})$.
Then the basis transformation is expressed as:

\begin{equation}
|u_m\rangle = \sum_\mathbf{q} |\mathbf{q}\rangle \langle
\mathbf{q}|u_m\rangle
\end{equation}

Here $\langle \mathbf{q}|u_m\rangle$ is simply the
$\mathbf{q}^{\mathrm{th}}$ Fourier component of the function
$u_m(\mathbf{r}_{\parallel})$\cite{Book_Cohen_Mec_Q}:

\begin{equation}\label{Fourier_trans}
\langle \mathbf{q}|u_m\rangle = \frac{1}{S} \iint
e^{i\mathrm{q}\mathbf{r}_{\parallel}} u_m(\mathbf{r}_{\parallel})
d^2 \mathbf{r}_{\parallel}
\end{equation}

We have, accordingly, the transformation law for the $B$-operators
\pref{B-op}:

\begin{equation}\label{B_m_transf}
B^\dagger_{\alpha m} = \sum_\mathbf{q} \langle
\mathbf{q}|u_m\rangle B^\dagger_{\alpha \mathbf{q}}
\end{equation}

Since the same transformation law applies to the bosonic operators
$b^\dagger_{\alpha \mathbf{q}}$, we can readily express the
expansion of the polarization density in the new basis:

\begin{eqnarray}\label{P_zbox}
\hat{P}_z(\mathbf{r}) = \frac{\hbar e}{2S m^\ast} \sum_{\alpha, m}
\frac{\sqrt{\Delta N_{\alpha}}}{\omega_{\alpha}}
\xi_{\alpha}(z)u_{m}(\mathbf{r}_{\parallel})[b^\dagger_{\alpha
m}+b_{\alpha m}]
\end{eqnarray}

The interaction Hamiltonian is also readily expressed. Details are
provided in Appendix \ref{AppTransfo}, and the final expression
for the plasma Hamiltonian of 0D microcavities is:

\begin{eqnarray}\label{Hplasma_m}
\hat{H} =  \sum_{\alpha, m}\hbar \widetilde{\omega}_\alpha
p^\dagger_{\alpha m}p_{\alpha m} +\sum_{m}\hbar
\omega_{c m}(a_m^\dagger a_m + 1/2)\nonumber\\
+i\sum_{\alpha, m} \frac{\hbar
\omega_{P\alpha}}{2}\sqrt{\frac{\omega_{c
m}}{\widetilde{\omega}_\alpha}f^o_{\alpha}f^w_{\alpha}}(a^\dagger_m-a_m)(p^\dagger_{\alpha
m}+p_{\alpha m})\nonumber\\
+\sum_{\alpha \neq \beta, m} \frac{\hbar \omega_{P\alpha}
\omega_{P\beta}}{4 \sqrt{\widetilde{\omega}_{\alpha}
\widetilde{\omega}_\beta}} C_{\alpha, \beta} (p^\dagger_{\alpha
m}+p_{\alpha m})(p^\dagger_{\beta m}+p_{\beta m})
\end{eqnarray}

The above Hamiltonian \pref{Hplasma_m} is formally similar to the
plasma Hamiltonian \pref{Hplasma_q}. The difference arises from
the modified spectrum of electromagnetic modes, which is no longer
continuum but discrete. The discrete spectrum allows to study
the simplest possible light-matter interacting system, where only
the lowest intersubband transition ($\alpha =2,1$) interacts
resonantly with a single microcavity mode that is sufficiently far
from the others, say the fundamental mode $m=0$. Then the plasma
Hamiltonian for this  "polariton dot" system is reduced to
\cite{Todorov_PRL2010}:

\begin{eqnarray}
\hat{H} = \hbar \widetilde{\omega}_{21} p^\dagger_{21}p_{21} +
\hbar \omega_{c}(a^\dagger a + 1/2) \nonumber\\+i\frac{\hbar
\omega_{P21}}{2}\sqrt{\frac{\omega_{c
}}{\widetilde{\omega}_{21}}f^o_{21}f^w_{21}}(a^\dagger-a)(p^\dagger_{21}+p_{21})
\end{eqnarray}

\begin{figure}
\includegraphics[scale=0.16]{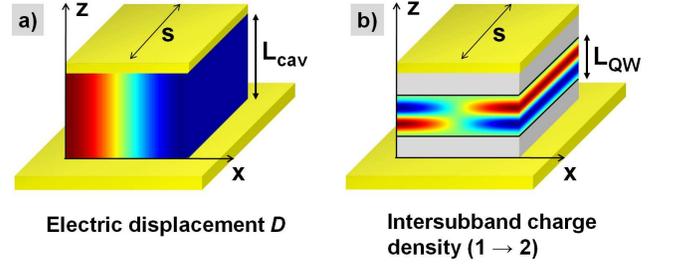}
\caption {a) Electrical displacement for the $(N=1, M=0)$ mode of
a patch microcavity with a patch size $s$. b) Intersubband charge
density for the fundamental transition of a quantum well with
thickness $L_{\mathrm{QW}}$ coupled with the $(N=1, M=0)$ mode of
the microcavity.} \label{Fig5}
\end{figure}

An example of a microcavity is provided by the square patch
resonators depicted in Refs. \cite{Todorov_Opex_2010,
Todorov_PRL2010}. In this case the resonances are labelled by two
lateral indexes $m=N,M$ and the corresponding lateral functions
are:

\begin{eqnarray}\label{um_zbox}
u_{N,M}(\mathbf{r}_{\parallel}) = \nonumber\\
\sqrt{2-\delta_{0N}}\sqrt{2-\delta_{0M}}\cos \Big{(} \frac{\pi N
x}{s}\Big{)}\cos \Big{(}\frac{\pi M y}{s}\Big{)}
\end{eqnarray}

Here $s$ is the size of the square. The fundamental mode is
actually twofold degenerate $(N=1, M=0)$,$(N=0, M=1)$, however
only one of the two modes can be selected in experiments with
polarized light. The mode $(N=1, M=0)$ is depicted in Figure
\ref{Fig5}(a).

Generally, in the case of a resonant light excitation, only the
spatial modes that correspond to the specific microcavity
resonance will be excited. That means that the charge distribution
will vibrate into well defined spatial modes, imposed by the
microcavity. The charge density $\rho (\mathbf{r})$ can be
obtained from the divergence of the polarization matrix element
\pref{P_zbox}, $\rho (\mathbf{r}) =\partial
P_z(\mathbf{r})/\partial z$:

\begin{equation}
\rho_{\lambda, \mu, m}(\mathbf{r}) = \frac{e\sqrt{\Delta
N_{\lambda \mu}}}{S} \phi_\lambda (z) \phi_\mu (z)
 u_{m}(\mathbf{r}_{\parallel})
\end{equation}

In the case of a square infinite quantum well of thickness
$L_{\mathrm{QW}}$ the wavefunctions are $\phi_\lambda (z) =
\sqrt{2/L_{\mathrm{QW}}}\sin (\lambda \pi z/L_{\mathrm{QW}})$. The
charge distribution which corresponds to the first intersubband
excitation $1 \rightarrow 2$ coupled with the fundamental cavity
mode $N=1, M=0$ has been plotted in Figure \ref{Fig5}(b).
This example shows how the microcavity allows to control the
spatial properties of the bright state.

\subsection{Back transformation in the Coulomb gauge}\label{min_coupling_gauge}

The light-matter coupling between microcavities and intersubband
transitions have been studied, so far, exclusively in the Coulomb
gauge \cite{Ciuti_PhysRevB_2005}. Usually, the case of a single
intersubband transition coupled with a continuum of guided modes
$\omega_{c\mathbf{q}}$ has been considered. These theoretical
studies pointed out the possibility to obtain the "ultra-strong"
coupling regime, where the light-matter coupling constant becomes
comparable to the frequency of the intersubband transition
$\omega_{21}$ \cite{Ciuti_PhysRevB_2005}. In this case the full
quantum Hamiltonian, including the anti-resonant terms and the
quadratic vector potential term $\mathbf{\hat{A}^2}$ must be taken
into account in order to describe correctly the system
\cite{Ciuti_PhysRevB_2005}.

In this section we establish a link between the previous studies
in the Coulomb gauge and the description in the dipole gauge
developed here. We shall consider the case of a single
intersubband transition ($\lambda=2, \mu=1$), coupled with the
$\mathrm{TM}_0$ mode. We then perform a unitary transformation to
the Hamiltonian in order to obtain its expression in the Coulomb
gage. The transformed Hamiltonian will have identical eigenvalues
as the original one, but will be expressed in terms of the vector
and scalar potentials $\mathbf{A}$ and $V$.

In this section, it is convenient to keep the individual subband
indexes $2,1$. We return to the form of the Hamiltonian with a
single intersubband transition before the Bogoliubov
transformation leading to the depolarization shift:

\begin{eqnarray}\label{H_12full}
\hat{H} = \sum_\mathbf{q}\hbar \omega_{21} b^\dagger_{\mathbf{q}}
b_{\mathbf{q}} +\sum_{\mathbf{q}}\hbar
\omega_{c \mathbf{q}}(a_\mathbf{q}^\dagger a_\mathbf{q} + 1/2)\nonumber\\
+i\sum_{\mathbf{q}} \frac{\hbar \omega_P}{2}\sqrt{\frac{\omega_{c
\mathbf{q}}}{\omega_{21}}f_{21}^o
f_{21}^w}(a^\dagger_\mathbf{q}-a_{-\mathbf{q}})(b^\dagger_{-\mathbf{q}}+b_{\mathbf{q}})
\nonumber\\
+\frac{\hbar
\omega_P^2}{4\omega_{21}}(b^\dagger_{\mathbf{q}}+b_{-\mathbf{q}})(b^\dagger_{-\mathbf{q}}+b_{\mathbf{q}})
\end{eqnarray}

(The subscripts "$21$" have been dropped for operators). The
vector potential for the $\mathrm{TM}_0$ mode is:

\begin{equation}\label{A-pot}
\mathbf{\hat{A}}(\mathbf{r}) = \mathbf{e}_z \sum_\mathbf{q}
\sqrt{\frac{\hbar}{2\varepsilon \varepsilon_0 S L_{\mathrm{cav}}
\omega_{c\mathbf{q}}}}e^{i\mathbf{q}\mathbf{r}_{\parallel}}
(a_\mathbf{q}+a_\mathbf{-q}^\dagger)
\end{equation}

We use \pref{A-pot} to express the inverse PZW unitary
transformation \cite{Babiker_Loudon_1983}:

\begin{equation}
T = \exp\Big{(} -\frac{i}{\hbar} \int
\mathbf{\hat{A}}(\mathbf{r})\cdot\mathbf{\hat{P}}(\mathbf{r})d^3
\mathbf{r} \Big{)}
\end{equation}

It writes in the case of a single intersubband transition:

\begin{eqnarray}\label{Inverse-T}
T = \exp \Big{(} -i \sum_\mathbf{q}
\chi_\mathbf{q}(a_\mathbf{q}^\dagger +
a_\mathbf{-q})(b_\mathbf{-q}^\dagger + b_\mathbf{q}) \Big{)} \\
\chi_\mathbf{q} = \frac{1}{2}\sqrt{\frac{\omega_P^2 f_{21}^o
f_{21}^w}{\omega_{21}\omega_{c\mathbf{q}}}}
\end{eqnarray}

Recalling the parity of the photon dispersion
$\omega_{c\mathbf{q}}=\omega_{c\mathbf{-q}}$ the following
transformation laws of the bosonic operators are obtained:

\begin{eqnarray}
T^+b_\mathbf{q} T = b_\mathbf{q}
-i\chi_\mathbf{q}(a_\mathbf{-q}^\dagger + a_\mathbf{q}) \\
T^+a_\mathbf{q} T = a_\mathbf{q}
-i\chi_\mathbf{q}(b_\mathbf{-q}^\dagger + b_\mathbf{q})
\end{eqnarray}

With these relations and their hermitian conjugates, the
transformed Hamiltonian is obtained to be:

\begin{eqnarray}\label{H_pA}
T^+\hat{H}T = \sum_\mathbf{q}\hbar \omega_{21}
b^\dagger_{\mathbf{q}} b_{\mathbf{q}} +\sum_{\mathbf{q}}\hbar
\omega_{c \mathbf{q}}(a_\mathbf{q}^\dagger a_\mathbf{q} + 1/2)\nonumber\\
+i\sum_{\mathbf{q}}\hbar \bar{\Omega}_\mathbf{q}
(a^\dagger_\mathbf{q}+a_{-\mathbf{q}})(b_{\mathbf{q}}-b^\dagger_{-\mathbf{q}})
\nonumber\\
+\sum_{\mathbf{q}} \frac{\hbar
\bar{\Omega}_\mathbf{q}^2}{\omega_{12}}(a^\dagger_{\mathbf{q}}+a_{-\mathbf{q}})
(a^\dagger_{-\mathbf{q}}+a_{\mathbf{q}})
\nonumber\\
+(1-f_{21}^o f_{21}^w)\sum_{\mathbf{q}} \frac{\hbar
\omega_P^2}{4\omega_{21}}(b^\dagger_{\mathbf{q}}+b_{-\mathbf{q}})(b^\dagger_{-\mathbf{q}}+b_{\mathbf{q}})
\end{eqnarray}

Here $\bar{\Omega}_\mathbf{q}$ is the light-matter coupling
constant in the minimal coupling gauge:

\begin{equation}\label{Rabi_Ap}
\bar{\Omega}_\mathbf{q} = \omega_{21}\chi_\mathbf{q}
=\frac{\omega_P}{2}\sqrt{\frac{\omega_{21}
}{\omega_{c\mathbf{q}}}f_{21}^o f_{21}^w}
\end{equation}

The first four terms of \pref{H_pA}, together with \pref{Rabi_Ap}
provide the Hopfield-like Hamiltonian used so for the theoretical
study of intersubband polaritons
\cite{Hopfield_1958,Ciuti_PhysRevB_2005}. Namely, the presence of
the $\hat{\mathbf{A}}^2$ term and the anti-resonant terms leads to
the "ultra-strong" coupling regime \cite{Ciuti_PhysRevB_2005}. The
later has been defined as the situation where the coupling term
$\bar{\Omega}_\mathbf{q}$, taken at resonance $\omega_{21} =
\omega_{c\mathbf{q}}$ becomes comparable with the energy of the
intersubband transition $\omega_{21}$:

\begin{equation}
\bar{\Omega}_\mathbf{q}(\omega_{21} = \omega_{c\mathbf{q}})
=\frac{\omega_P}{2}\sqrt{f_{21}^o f_{21}^w} \approx \omega_{21}
\end{equation}

Since from \pref{H_pA} and \pref{Rabi_Ap} the $\hat{\mathbf{A}}^2$
term is proportional to square of the plasma frequency
$\omega_P^2$, we see that the ultra-strong coupling regime, as
defined in Ref. \cite{Ciuti_PhysRevB_2005} is obtained in systems
featuring high photonic confinement factor $f_{21}^w$, and high
plasma frequency $\omega_P$, i.e. high electronic densities.
However, for high electronic densities the Coulomb interaction
bring important dynamical corrections \cite{Keeling_2007}. The
latter are already present in the dipole gauge, which naturally
includes the collective excitations of the electron gas, as
described in section \ref{Sec_PlasmaH}. In the Hamiltonian
\pref{H_pA} expressed in the Coulomb gauge, these correction
actually arise from the last term. In Appendix \ref{AppCoulomb} we
show that this term can indeed be cast in a form of a
long-wavelength limit of the Coulomb potential.

The Coulomb correction in the Hamiltonian \pref{H_pA} contains an
interesting new element, which is the geometrical prefactor
$(1-f_{21}^o f_{21}^w)$. This prefactor is equal to one in the
case of a very large cavity ($f_{21}^w \rightarrow 0$). This case
of vanishing photon confinement corresponds, for instance, to the
multi-pass waveguides employed for the absorption measurements
where the depolarization shift is observed. However, the
confinement factor $f_{21}^of_{21}^w$ becomes an important
correction in the case of micro-cavities with filling factors
$f_{21}^w$ close to unity, such as the double metal microcavities
\cite{Todorov_PRL2010, geiser_2010, PJouy_2011}. We interpret the
factor $-f_{21}^o f_{21}^w$ as an image contribution to the
Coulomb interaction due to the boundary conditions of the electric
field at the microcavity walls. In other words, it can be seen as
a local field correction due the partial screening of the
microcavity field by the oscillating intersubband charges. Indeed,
both the displacement field $\mathbf{\hat{D}}_z(\mathbf{r})$ and
the polarization field $\mathbf{\hat{P}}_z(\mathbf{r})$ that we
used to construct the PZW Hamiltonian satisfy the boundary
conditions on the cavity walls, which are transported to 
the Coulomb correction through the unitary transformation \pref{Inverse-T}.

Note that the the unitary transformation \pref{Inverse-T} holds
only for a truncated Hilbert space, where we retained only the
first intersubband transition and the fundamental waveguide
mode. The full gauge equivalence is established correctly only if
the complete set of quantum transitions and electromagnetic modes
of the system are accounted for \cite{Bassani_1977}. This issue
will be discussed elsewhere.

We can combine the last term in the Hamiltonian \pref{H_pA} with
the first one, and perform a Bogoliubov diagonalization just like
in section \ref{Sec_PlasmaH}. As a result, we obtain a
renormalized intersubband frequency with an effective
depolarization shift, which takes into account the local field
corrections, contained in the factor $f_{21}^o f_{21}^w$:

\begin{equation}\label{w21bar}
\bar{\omega}_{21} = \sqrt{\omega^2_{21}+\omega_P^2(1-f_{21}^o
f_{21}^w)}
\end{equation}

The local field factor $1-f_{21}^o f_{21}^w$ yields an
effective plasma frequency $\omega_P\sqrt{1-f_{21}^o f_{21}^w}$.
We can then rewrite the full Hamiltonian in the Coulomb gauge
\pref{H_pA} using equation \pref{w21bar} and the corresponding
polarization operators. The result is exactly the Hamiltonian of
Ref. \cite{Ciuti_PhysRevB_2005}, used for the study of the
ultra-strong coupling regime, where the bare intersubband
frequency has been replaced by the effective frequency
$\bar{\omega}_{21}$:

\begin{eqnarray}\label{H_pA_bis}
T^+\hat{H}T = \sum_\mathbf{q}\hbar \bar{\omega}_{21}
b^\dagger_{\mathbf{q}} b_{\mathbf{q}} +\sum_{\mathbf{q}}\hbar
\omega_{c \mathbf{q}}(a_\mathbf{q}^\dagger a_\mathbf{q} + 1/2)\nonumber\\
+i\sum_{\mathbf{q}}\hbar \bar{\Omega}_\mathbf{q}
(a^\dagger_\mathbf{q}+a_{-\mathbf{q}})(b_{\mathbf{q}}-b^\dagger_{-\mathbf{q}})
\nonumber\\
+\sum_{\mathbf{q}} \frac{\hbar
\bar{\Omega}_\mathbf{q}^2}{\bar{\omega}_{21}}(a^\dagger_{\mathbf{q}}+a_{-\mathbf{q}})
(a^\dagger_{-\mathbf{q}}+a_{\mathbf{q}})
\end{eqnarray}

Note that now $b_{\mathbf{q}}$ describes the bosonic operator
after the Bogoliubov transformation, and the frequency
$\bar{\omega}_{21}$ should be used in the definition of
$\bar{\Omega}_\mathbf{q}$. This result validates the studies
performed in the Coulomb gauge \cite{Anappara_2009a}, in the limit
of microcavities with large filling factors, where
$\bar{\omega}_{21} \approx \omega_{21}$. However, contrary to Refs. \cite{Ciuti_PhysRevB_2005} and \cite{Anappara_2009a}, the effective Hamiltonian 
\pref{H_pA_bis} includes also the correct limit of the weak coupling regime, obtained
for vanishing overlap with the cavity mode $f_{21}^w \rightarrow 0$.
In this case from expression \pref{w21bar} we recover the renormalized transition 
frequency $\widetilde{\omega}_{21}$, whereas the Hamiltonians in Refs. \cite{Ciuti_PhysRevB_2005} and \cite{Anappara_2009a} would predict only the 
bare intersubband spacing $\omega_{21}$.

\section{Properties of the polariton states}\label{Part_III}

\subsection{Polariton dispersion}\label{Sec_Dispersion}

We now analyse coupled light-matter polariton states arising from the
dipolar Hamiltonian. To simplify, we ignore the coupling between plasmons on
different subbands. This is equivalent to consider a single
intersubband transition $\omega_\alpha$, not necessarily the
fundamental one, in interaction with the $\mathrm{TM}_0$ mode. All
the results that will be stated remain also valid for 0D
resonators. The corresponding plasma Hamiltonian is:

\begin{eqnarray}\label{H_12}
\hat{H} = \sum_\mathbf{q}\hbar \widetilde{\omega}_{\alpha}
p^\dagger_{\mathbf{q}} p_{\mathbf{q}} +\sum_{\mathbf{q}}\hbar
\omega_{c \mathbf{q}}(a_\mathbf{q}^\dagger a_\mathbf{q} + 1/2)\nonumber\\
+i\sum_{\mathbf{q}} \frac{\hbar \omega_{P
\alpha}}{2}\sqrt{\frac{\omega_{c
\mathbf{q}}}{\widetilde{\omega}_{\alpha}}f_{\alpha}^o
f_{\alpha}^w}(a^\dagger_\mathbf{q}-a_{-\mathbf{q}})(p^\dagger_{-\mathbf{q}}+p_{\mathbf{q}})
\end{eqnarray}

This Hamiltonian is very similar to a Dicke model
\cite{Dicke_1954}. However, the coupling coefficient is
proportional to $\omega_P/\sqrt{\widetilde{\omega}_{\alpha}}$ and
has a non-linear dependence on $\omega_P$ because of the formula
of the depolarization shift
$\widetilde{\omega}_{\alpha}=\sqrt{\omega_{\alpha}^2
+\omega_P^2}$. We show further that this non-linearity leads to
the no-go theorem for quantum well systems, and the Hamiltonian
\pref{H_12} does not allow a quantum phase transition
\cite{Nataf_Ciuti_2010}. The Hamiltonian \pref{H_12} can be
diagonalized exactly by introducing the polariton operator:

\begin{equation}
\Pi_\mathbf{q} = x_\mathbf{q}a_\mathbf{q}  +
y_\mathbf{q}a_\mathbf{-q}^\dagger + z_\mathbf{q}p_{\mathbf{q}} +
t_\mathbf{q}p^\dagger_{-\mathbf{q}}
\end{equation}

The Hopfield coefficients introduced here satisfy the
normalization condition:

\begin{equation}
|x_\mathbf{q}|^2 - |y_\mathbf{q}|^2 + |z_\mathbf{q}|^2 -
|t_\mathbf{q}|^2 = 1
\end{equation}

The Hopfield-Bogoliubov determinant corresponding to the equation
$[\hat{H}, \Pi_\mathbf{q}]=\hbar \omega_\mathbf{q} \Pi_\mathbf{q}$
is then:

\begin{eqnarray}\label{Dicke_Determinant}
\left\|
\begin{array}{cccc}
\omega_{c\mathbf{q}} - \omega_\mathbf{q} & 0 & i\Omega_\mathbf{q} &  i\Omega_\mathbf{q} \\
0 & -\omega_{c\mathbf{q}} - \omega_\mathbf{q} &  i\Omega_\mathbf{q} &  i\Omega_\mathbf{q} \\
-i\Omega_\mathbf{q} &  i\Omega_\mathbf{q} & \widetilde{\omega}_{\alpha} - \omega_\mathbf{q} & 0 \\
i\Omega_\mathbf{q} & -i\Omega_\mathbf{q} & 0 &
-\widetilde{\omega}_{\alpha} - \omega_\mathbf{q}
\end{array} \right\|
\end{eqnarray}

with $\Omega_\mathbf{q}$ the light-matter coupling constant from
\pref{Coupling_Const_Dipgauge} (the subscript $\alpha$ has been
dropped). Zeroing the determinant \pref{Dicke_Determinant}
provides the eigenvalue equation:

\begin{equation}\label{EigenEq1}
(\omega_\mathbf{q}^2 -
\widetilde{\omega}_{\alpha}^2)(\omega_\mathbf{q}^2 -
\omega_{c\mathbf{q}}^2) = f_{\alpha}^o f_{\alpha}^w
\omega_{P\alpha}^2 \omega_{c\mathbf{q}}^2
\end{equation}

This biquadratic equation can be solved analytically, the two real
solutions, $\omega_{\mathbf{q},+}$ and $\omega_{\mathbf{q},-}$,
being the frequencies of the two coupled states. The Hopfield
coefficients are also readily expressed in closed form. For
instance, we can define a "photonic" part $h_p=|x_\mathbf{q}|^2 -
|y_\mathbf{q}|^2$ and an "electronic" part $h_e = |z_\mathbf{q}|^2
- |t_\mathbf{q}|^2$ linked by the relation $h_p+h_e=1$. For the
photonic part we obtain the expressions:

\begin{equation}\label{Hop_coeffs}
h_{p,+} =
\frac{\omega^2_{\mathbf{q},+}-\widetilde{\omega}_{\alpha}^2}{\omega^2_{\mathbf{q},+}-\omega^2_{\mathbf{q},-}},
\phantom{Q} h_{p,-} =
\frac{\widetilde{\omega}_{\alpha}^2-\omega^2_{\mathbf{q},-}}{\omega^2_{\mathbf{q},+}-\omega^2_{\mathbf{q},-}}
\end{equation}

Note that we have necessarily $h_{p,+}+h_{p,-}=1$ (and therefore
$h_{e,+}+h_{e,-}=1$).

\begin{figure}
\includegraphics[scale=0.12]{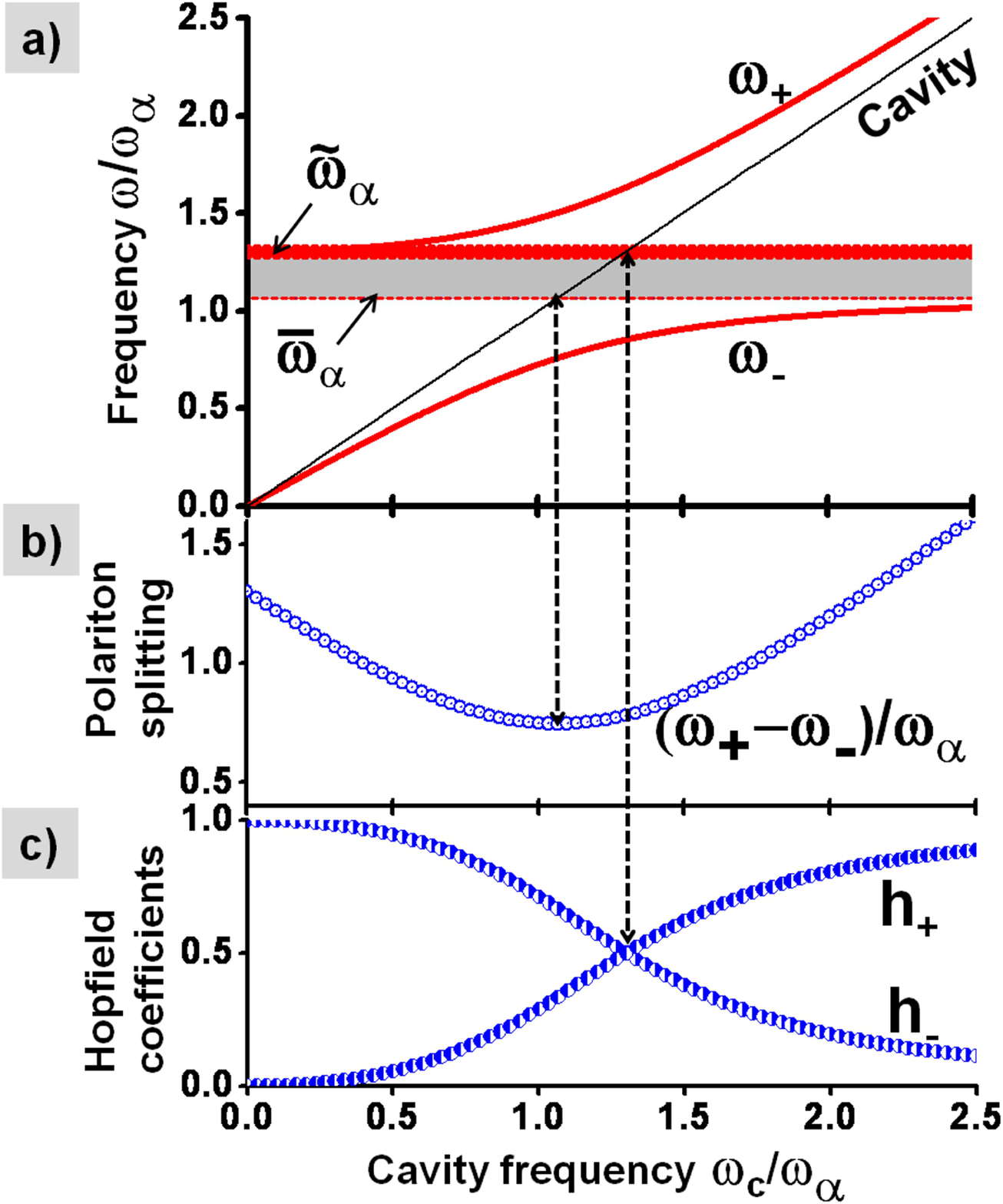}
\caption {(a) Polariton dispersion, normalized at the bare
intersubband transition $\omega_{\alpha}$ (b) Splitting of the two
polariton states. (c) Hopfield coefficients. In these figures, the
subscript $\mathbf{q}$ has been dropped, for clarity. The
numerical values used for solving equation \pref{EigenEq1} are
$\omega_P = 0.83\omega_{\alpha}$, $f^o_{\alpha}=1$ and
$f^w_{\alpha}=0.8$.} \label{Fig6}
\end{figure}

The polariton frequencies  $\omega_{\mathbf{q},\pm}$, as well as
the electronic Hopfield coefficients $h_{e,\pm}$ have been plotted
as a function of the cavity frequency $\omega_{c\mathbf{q}}$ on
Figure \ref{Fig6}(a)(c). For this illustration we have used
the numerical values $\omega_P = 0.83\omega_{\alpha}$,
$f^o_{\alpha}=1$ and $f^w_{\alpha}=0.8$. As seen from Figure
\ref{Fig6}(a) the polarion dispersion features a gap. The
upper edge of the gap, obtained at $\omega_{c\mathbf{q}}=0$ is the
frequency of the intersubband plasmon
$\widetilde{\omega}_{\alpha}$. The lower edge is easily estimated
from equation \pref{EigenEq1} to be:

\begin{equation}\label{Low_gapEdge}
\omega_{\mathbf{q},-}|_{\omega_{c\mathbf{q}} \rightarrow \infty} =
\sqrt{\omega^2_{\alpha}+\omega_{P\alpha}^2(1-f_{\alpha}^o
f_{\alpha}^w)} = \bar{\omega}_{\alpha}
\end{equation}

In the next section we show that $f_{\alpha}^o f_{\alpha}^w<1$,
therefore the lower gap edge always appears at a frequency higher
than the bare intersubband frequency $\omega_{\alpha}$. The
impossibility for the light to propagate at the gap energies can
be explained by the destructive interference between the
microcavity electromagnetic field and the local field created by
the collective electronic oscillations.

The most important aspect of the strong-light matter coupling
regime is the mixing between the electronic and photonic degrees
of freedom. This mixing is quantified by the Hopfield coefficients
\pref{Hop_coeffs}. Namely, the coupled system features maximum
mixing when the photonic part of the Hopfield coefficients equals
the electronic part: $h_{e,\pm}=h_{p,\pm}=1/2$. From equation
\pref{Hop_coeffs} we readily obtain that this is satisfied when
the cavity is tuned into resonance with the intersubband plasmon:

\begin{equation}
h_{e,\pm}=h_{p,\pm} \Leftrightarrow \omega_{c\mathbf{q}} =
\widetilde{\omega}_{\alpha}
\end{equation}

This has been illustrated in Figure \ref{Fig6}(c). In the well
known Jaynes-Cummings model \cite{Book_Fox_QuantOpt} the maximum
mixing also corresponds to the point of minimal splitting
$\omega_{\mathbf{q},+} - \omega_{\mathbf{q},-} $ between the two
polariton states. However, this is not true in the general case.
As shown in Appendix \ref{AppPropDisp}, the minimum splitting
occurs when the cavity is resonant with the lower gap edge
frequency (equation \pref{Low_gapEdge}):

\begin{equation}
\frac{\mathrm{d}(\omega_{\mathbf{q},+} -
\omega_{\mathbf{q},-})}{\mathrm{d}\omega_{c\mathbf{q}}}=0
\Leftrightarrow \omega_{c\mathbf{q}} = \bar{\omega}_\alpha
\end{equation}

This is also illustrated in Figure \ref{Fig6}(b). Moreover, in
Appendix \ref{AppPropDisp} we show that the minimal splitting can
be computed exactly, and the results is:

\begin{figure}
\includegraphics[scale=0.130]{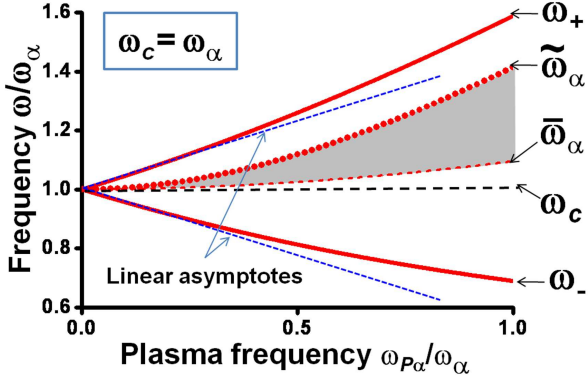}
\caption {Polariton frequencies as a function of the plasma
frequency for a cavity resonant with the bare intersubband
transition $\omega_{c\mathbf{q}}=\omega_{\alpha}$. The grey area
indicates the gap. For this figure we have used $f^o_{\alpha}=1$
and $f^w_{\alpha}=0.8$. The linear asymptotes correspond to the
Jaynes-Cummings model.} \label{Fig7}
\end{figure}

\begin{equation}\label{Rabi1}
\min(\omega_{\mathbf{q},+} - \omega_{\mathbf{q},-}) =
\sqrt{f_{\alpha}^o f_{\alpha}^w} \omega_P = 2\Omega_R
\end{equation}

We recognize the quantity $2\Omega_R$ to be the Rabi splitting, as
known from the usual definition \cite{Andreani_2003}:

\begin{equation}\label{Rabi2}
2\Omega_R = \sqrt{\frac{f_{\alpha}^o e^2 (\Delta
N_{\alpha}/S)}{\varepsilon \varepsilon_0 m^\ast L_{\mathrm{cav}}}}
\end{equation}

Note that this is an exact result, that is valid for all orders of
light-matter interaction. More general result for an arbitrary
cavity can be obtained with the expressions
\pref{Coupling_Const_General} or \pref{General_fw}. The quantity
$2\Omega_R$ can therefore be used as an experimental measure of
the interaction strength, even in the ultra-strong coupling
regime.

We see that both edges of the gap, $\bar{\omega}_\alpha$ and
$\widetilde{\omega}_{\alpha}$ play an important role in the
theory. The frequency $\widetilde{\omega}_{\alpha}$ defines the
point of maximum quantum mixing, whereas $\bar{\omega}_\alpha$
defines the point of minimum splitting $2\Omega_R$. However, we
can observe from the plot on Figure \ref{Fig6}(b) that the
splitting around the point $\omega_{c\mathbf{q}} =
\widetilde{\omega}_{\alpha}$ is not very different from
$2\Omega_R$. This feature is a characteristic of the ultra-strong
coupling regime, where we can no longer define a strict
resonance condition for the optimal coupling point. Indeed, the
system will feature almost identical coupling energy $2\hbar
\Omega_R$ if the cavity resonates with any frequency in the band
between $\bar{\omega}_\alpha$ and $\widetilde{\omega}_{\alpha}$.
The three characteristic frequencies
$\widetilde{\omega}_{\alpha}$, $\bar{\omega}_{\alpha}$ and
$2\Omega_R$ are liked through the simple relation:

\begin{equation}
\widetilde{\omega}^2_{\alpha} =
\bar{\omega}^2_{\alpha}+4\Omega^2_R
\end{equation}

As pointed out in Ref. \cite{Ciuti_PhysRevB_2005}, for a very
large interaction strength, the effects of the quadratic and
anti-resonant terms in the interaction Hamiltonian become
important. These effects are manifested with the non-linear
behavior of the polariton frequencies $\omega_{\mathbf{q},\pm}$ as
a function of the Rabi splitting $2\Omega_R$
\cite{Ciuti_PhysRevB_2005}. Since the light-matter interaction
strength scales as the plasma frequency $\omega_P$ (equations
\pref{Rabi1} and \pref{Rabi2}) we have studied these effects as a
function of $\omega_P$, as shown in Figure \ref{Fig7}.
Experimentally, $\omega_P$ can be varied either through the
temperature of the system, or by applying a gate voltage
\cite{Anappara_2006}. In both cases one controls the subband
population difference $\Delta N_\alpha$. For the plot of Figure
\ref{Fig7} we have chosen a cavity that is resonant
with the bare intersubband transition when $\omega_P=0$, i.e.
$\omega_{c\mathbf{q}}=\omega_{\alpha}$. Moreover, in this example,
we neglect the Hartree correction of the bare intersubband frequency as
the number of charges in the system progressively increases. This assumption
is true for a sufficiently thin square quantum well.
Because both frequencies $\bar{\omega}_\alpha$ and
$\widetilde{\omega}_{\alpha}$ increase
with $\omega_P$ due to the depolarization effect, the system gets
blue-shifted from the cavity mode, however the blue shift is much
smaller for $\bar{\omega}_{\alpha}$. The Jaynes-Cummings model
corresponds to the linear asymptotes of the polariton branches at
low $\omega_P$. Figure \ref{Fig7} clearly shows that
this model ceases to be valid for high electronic densities.
Moreover, the magnitude of the polariton gap increases as the
polariton branches depart from the linear asymptotes. Therefore
the measurement of the gap can be considered as a direct
spectroscopic evidence of the ultra-strong coupling regime
\cite{Todorov_PRL2010, PJouy_2011}.

\begin{figure}
\includegraphics[scale=0.140]{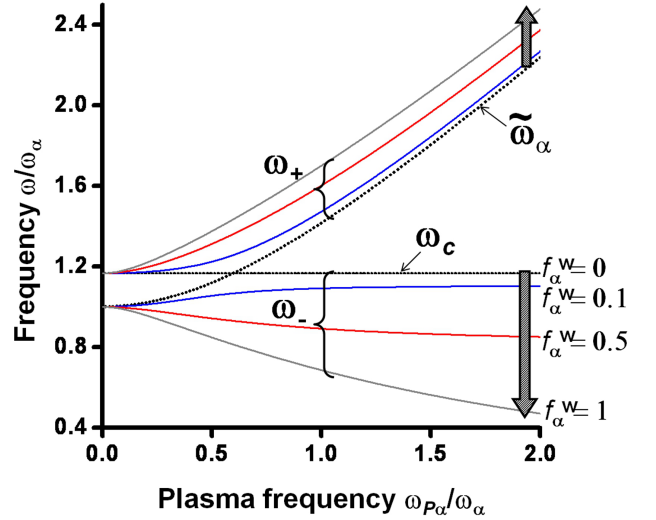}
\caption {Illustration of the polariton frequencies for cavities
with increasing filling factor $f^w_{\alpha}$. For this
illustration, the cavity is slightly red-shifted as respect to the
bare cavity frequency $\omega_\alpha$.} \label{Fig8}
\end{figure}

Along with the charge density, the second ingredient of the light
-matter interaction is the overlap factor $f^w_\alpha$ between the
photonic mode and the intersubband current. As evident from the
expression of the Hamiltonian in the Coulomb gauge \pref{H_pA},
this factor controls not only the intensity of the light-mater
coupling (linear term in equation \pref{H_pA}), but also the
relative weigh between the longitudinal Coulomb corrections (last
term in \pref{H_pA}) and the transverse corrections contained in
the $\mathbf{A}^2$ term.

The influence of the overlap factor $f^w_\alpha$ on the polariton
dispersion is illustrated in Figure \ref{Fig8}. In this plot,
the cavity frequency $\omega_{c\mathbf{q}}$ is slightly
blue-shifted as respect to the bare intersubband transition
$\omega_{\alpha}$. When $f^w_{\alpha}=0$, we recover two uncoupled
oscillators at frequencies $\omega_{c\mathbf{q}}$ and
$\widetilde{\omega}_{\alpha}$ as expected from equation
\pref{EigenEq1}. For small values of $f^w_\alpha$, there is a
small splitting that appears when the cavity is tuned with the
intersubband plasmon, $\omega_{c\mathbf{q}}=\widetilde{\omega}_{\alpha}$.
In this case the $\mathbf{A}^2$ term of the Hamiltonian \pref{H_pA}, that
scales as $(f^w_\alpha)^2$ is negligible. The polariton gap is
also negligible, since $\widetilde{\omega}_{\alpha} \approx
\bar{\omega}_{\alpha}$, and the system is described by
Jaynes-Cummings model with a resonant frequency
$\widetilde{\omega}_{\alpha}$ renormalized by the Coulomb
interactions.

On the contrary, for filling factors close to one, $f^w_\alpha
\approx 1$ the Coulomb correction in the Hamiltonian \pref{H_pA}
is negligible, and the weight of the quadratic term shifts to
$\mathbf{A}^2$. If the electronic density is sufficiently high,
the system enters the ultra-strong coupling regime as defined in
Ref. \cite{Ciuti_PhysRevB_2005}. The resonance condition
$\omega_{c\mathbf{q}}=\widetilde{\omega}_{\alpha}$ looses its
strict meaning, as the cavity can be resonant with any frequency
in the polariton gap.

\subsection{No-go theorem}\label{Sec_NoGo}

The "No-go" theorem states the impossibility of the lower
polariton state to acquire zero energy when the light-matter
interaction is increased \cite{Birula_1979}. This property is
related from the quadratic term in the interaction light-matter
Hamiltonian. For intersubband polaritons, the No-go theorem
becomes particulary clear in the dipole gauge, where, as shown
bellow, it stems from the depolarization effect.

\begin{figure}
\includegraphics[scale=0.17]{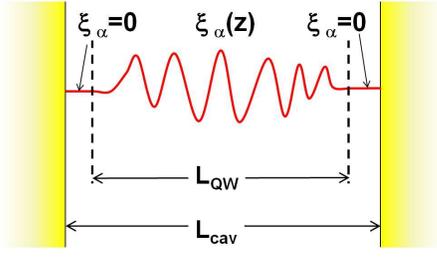}
\caption {Sketch of the intersubband current $\xi_{\alpha} (z)$
for a heterostructure in a cavity with thickness
$L_{\mathrm{cav}}$. We suppose that the current decays sufficiently fast,
so it can be considered as null close to the cavity boundaries. We
can then define a region of space with a thickness
$L_{\mathrm{QW}}<L_{\mathrm{cav}}$ containing the entire current density.}
\label{Fig9}
\end{figure}

The asymptotic value of the lower polariton frequency for very
high plasma frequencies $\omega_{P\alpha}\rightarrow \infty$ is
deduced from equation \pref{EigenEq1}:

\begin{equation}
\omega^2_-|_{\omega_{P\alpha} \rightarrow \infty} \rightarrow
\omega^2_{c\mathbf{q}} (1 - f_{\alpha}^w f_{\alpha}^o)
\end{equation}

Note that this limit does not contain the bare intersubband
frequency $\omega_\alpha$, and therefore is independent from the
eventual Hartree corrections to the heterostructure potential.
The No-go theorem for intersubband polariton is then equivalent to
the following strict inequality:

\begin{equation}
f_{\alpha}^w f_{\alpha}^o < 1
\end{equation}

This inequality can be proven by using the properties of the
intersubband current matrix element $\xi_\alpha (z)$
\pref{xi-density} which allows to express the overlap factor
$f_{\alpha}^w = L^{\alpha}_{\mathrm{eff}}/L_{\mathrm{cav}}$
through equation \pref{Leff}:

\begin{equation}\label{Recall_Leff}
f_{\alpha}^w = \frac{2m^\ast \omega_{\alpha}}{\hbar
L_{\mathrm{cav}} }\frac{1}{\int_{-\infty}^{\infty} \xi^2_\alpha
(z) dz}
\end{equation}

We suppose that wavefunctions of the bound states and their
derivatives decay sufficiently fast, so that $\xi_\alpha (z)$ is
zero close to the cavity boundaries, as illustrated in Figure
\ref{Fig9}. Then we can define a domain with a finite size
$L_{\mathrm{QW}} < L_{\mathrm{cav}}$, that we can arbitrarily call
"quantum well", such as all the the wavefunctions  and their
derivatives are zero outside this domain (Figure \ref{Fig9}). 
We chose the origin of the coordinates so that $0 \leqslant z
\leqslant L_{\mathrm{QW}}$ inside the "quantum well". We then
re-express the current-current integral in the non-local form:

\begin{eqnarray}\label{I_delta}
\int_{-\infty}^{+\infty} \xi_{\alpha} (z)^2 dz = \nonumber \\
 \iint_{-\infty}^{+\infty}  \xi_{\alpha} (z) \xi_{\alpha}(z') \delta(z-z')
 dzdz'
\end{eqnarray}

We can choose an arbitrary orthogonal basis of real functions
$\chi_n (z)$ on the segment $[0, L_{\mathrm{QW}}]$ in order to
span the delta-function:

\begin{equation}
\delta(z-z') = \sum_n \chi_n (z)\chi_n (z'),
\end{equation}

This basis is not necessarily the basis of envelope wavefunctions.
The expansion of the integral \pref{I_delta} on the basis $\chi_n
(z)$ is:

\begin{equation}\label{I_expansion}
\int_{-\infty}^{+\infty} \xi_{\alpha} (z)^2 dz = \sum_n  \Big{(}
\int_0^{L_{\mathrm{QW}}} \xi_{\alpha} (z)\chi_n (z) dz \Big{)}^2
\end{equation}

An evident choice for $\chi_n (z)$ is the Fourier basis on the
segment $[0, L_{\mathrm{QW}}]$:

\begin{eqnarray}
\chi_n (z) =
\begin{array}{cc}
\sqrt{\frac{2-\delta_{0,n}}{L_{\mathrm{QW}}}} \cos
\Big{(}\frac{2\pi n z}{L_{\mathrm{QW}}}\Big{)} & n \phantom{Q} \mathrm{even} \\
\sqrt{\frac{2}{L_{\mathrm{QW}}}}\sin\Big{(}\frac{2\pi n
z}{L_{\mathrm{QW}}}\Big{)} & n \phantom{Q} \mathrm{odd}
\end{array}
\end{eqnarray}

With this particular choice, the first basis function $\chi_0(z) =
\sqrt{1/L_{\mathrm{QW}}}$ is constant on the segment $[0,
L_{\mathrm{QW}}]$. Since all terms of equation \pref{I_expansion}
are positive, we have the Cauchy-Schwartz inequality for the
current-current overlap:

\begin{equation}
\int_{-\infty}^{+\infty} \xi_{\alpha} (z)^2 dz >
\frac{1}{L_{\mathrm{QW}}} \Big{(} \int_0^{L_{\mathrm{QW}}}
\xi_{\alpha} (z) dz \Big{)}^2
\end{equation}

The right hand side of the last equation can be expressed with the
oscillator strength according to the identities
\pref{current_identity} and \pref{ostrength}. Combining this
expression with the definition \pref{Recall_Leff}, we obtain the
following inequality:

\begin{equation}\label{Leff_inequality}
L^{\alpha}_{\mathrm{eff}} < \frac{L_{\mathrm{QW}}}{f_{\alpha}^o}
\end{equation}

This result shows that the higher the oscillator strength of an
intersubband transition, the more the corresponding current
density is localized in the space. This provides directly the
No-go theorem, since it leads to the inequality:

\begin{equation}
f_{\alpha}^w f_{\alpha}^o <
\frac{L_{\mathrm{QW}}}{L_{\mathrm{cav}}} < 1
\end{equation}

\begin{figure}
\includegraphics[scale=0.12]{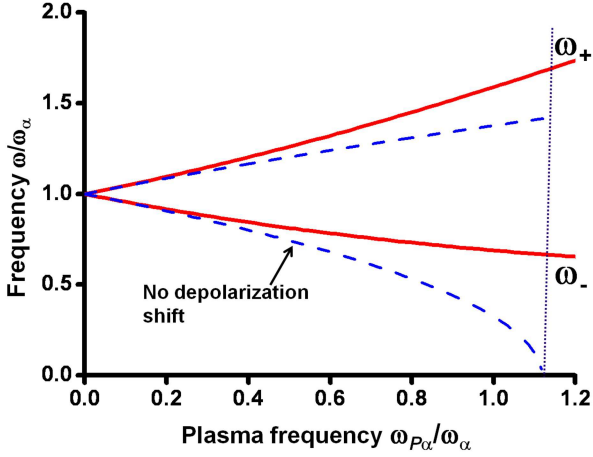}
\caption {Comparison of the polariton dispersion, at resonance
($\omega_{c\mathbf{q}}=\omega_\alpha$), with and without the
depolarization effect. For simplicity, the Hartree shift of
$\omega_\alpha$ is neglected in this illustration.} \label{Fig10}
\end{figure}

Therefore the No-go theorem is a consequence of the confinement of
the electronic plasma in a finite volume of space
\cite{Nataf_Ciuti_2010}. This confinement is also the origin of
the depolarization shift, which arises from the plasma energy of
the bound charges.

The inequality \pref{Leff_inequality} provides a lower bound for
the plasma energy of the system. The plasma energy is quantified
as the weight of the $\mathbf{\hat{P}}^2$ term in the plasma
Hamiltonian \pref{Hselfenergy}:

\begin{equation}
\hbar \frac{\omega_{P\alpha}^2}{4\omega_\alpha} = \frac{\hbar
e^2}{\varepsilon \varepsilon_0 m^\ast \omega_\alpha} \times
\frac{\Delta N_{\alpha}}{4S L^{\alpha}_{\mathrm{eff}}}
\end{equation}

Using the inequality \pref{Leff_inequality} and the definition of
the oscillator strength \pref{ostrength}, we obtain the following
inequality for the plasma energy:

\begin{equation}\label{omega_P_inequality}
\hbar \frac{\omega_{P\alpha}^2}{4\omega_\alpha} >
\frac{d_\alpha^2}{2 \varepsilon \varepsilon_0} \times \frac{\Delta
N_{\alpha}}{S L_{\mathrm{QW}}}
\end{equation}

Here $d_\alpha = e z_\alpha$ is the dipole moment of the
intersubband transition. The left side of
\pref{omega_P_inequality} corresponds exactly to the classical
self-interaction energy of $\Delta N_{\alpha}$ dipoles uniformly
distributed in a volume $SL_{\mathrm{QW}}$
\cite{Book_Cohen_Ph_At}. According to equation
\pref{omega_P_inequality} the quantum self-energy is higher than
the classical one, since the wavefunctions of the bound states
illustrated in Figure \ref{Fig2} provide stronger spatial
confinement of the charged particles.

The plasma energy of the charged particles localized in a finite
volume compensates for the energy decrease
of the lower polariton as the light-matter
coupling strength is increased. This is illustrated in Figure
\ref{Fig10}, where we have compared the polariton branches
$\omega_{\pm}$ as a function of $\omega_P$ with the solution of
equation \pref{EigenEq1} when the depolarization shift is
discarded, i.e. when $\widetilde{\omega}_\alpha$ is replaced with
$\omega_\alpha$. The cavity is chosen so that
$\omega_{c\mathbf{q}}=\omega_\alpha$, and the Hartree corrections
have been neglected for clarity. It is clearly seen that the
depolarization effect prevents the lower polariton energy to reach
zero, which is an equivalent statement of the No-go theorem for
intersubband transitions.

\subsection{Effective dielectric function}\label{DielFuncSection}

The dispersion relation \pref{EigenEq1} allows to obtain the
effective dielectric constant $\varepsilon_{\mathrm{eff}}(\omega)$
of the polaritonic medium. The latter can be defined through the
propagation equation:

\begin{equation}\label{deff_ep_eff}
\varepsilon_{\mathrm{eff}}(\omega_\mathbf{q})\frac{\omega_\mathbf{q}^2}{c^2}
= \mathbf{q}^2
\end{equation}

Since for the $\mathrm{TM}_0$ mode $\omega_{c\mathbf{q}}^2 = c^2
\mathbf{q}^2/\varepsilon$, we obtain from \pref{EigenEq1}:

\begin{equation}\label{ep_eff}
\frac{\varepsilon}{\varepsilon_{\mathrm{eff}}(\omega)} =
1+\frac{f_{\alpha}^o f_{\alpha}^w \omega_{P\alpha}^2}{\omega^2 -
\widetilde{\omega}_{\alpha}^2}
\end{equation}

Note that the zero of the dielectric constant
$\varepsilon_{\mathrm{eff}}(\omega)$ corresponds to the frequency
of the intersubband plasmon $\widetilde{\omega}_{\alpha}$, as
expected. The above expression can be recast in the form:

\begin{equation}\label{Qe_eff}
\frac{\varepsilon}{\varepsilon_\mathrm{eff}(\omega)} =
\frac{1-f_{\alpha}^o f_{\alpha}^w}{\varepsilon}+\frac{f_{\alpha}^o
f_{\alpha}^w}{\varepsilon_{\mathrm{QW}}(\omega)}
\end{equation}

where $\varepsilon_\mathrm{QW}(\omega)$ is the dielectric constant
of the heterostructure alone:

\begin{equation}
\varepsilon_\mathrm{QW}(\omega) =\varepsilon \Big{(}
1-\frac{\omega_{P\alpha}^2}{\omega^2 - \omega_{\alpha}^2} \Big{)}
\end{equation}

The result \pref{Qe_eff} is very similar to the effective
dielectric constant obtained by Zeluzny et Nalewajko
\cite{Zaluzny_Nalewajko_1999}. This result arises naturally from
the initial assumption that the heterostructure is much smaller
than the wavelength of light. However, in our case the geometrical
overlap factor $f_{\alpha}^o f_{\alpha}^w$ contains a quantum
correction due to the shape of the wavefunctions.

The model developed so far is purely Hamiltonian, and the
dissipation has not been included. The dissipation can be
introduced through a coupling
with a reservoir of harmonic oscillators. Such model has been
considered, for instance, in Refs. \cite{Huttner_Barnet_1992,
Dutra_Furuya_1998}. In particular, the method of Ref.
\cite{Dutra_Furuya_1998} leads to a dielectric function of the
form:

\begin{equation}
\varepsilon_\mathrm{QW}(\omega) =\varepsilon \Big{(}
1-\frac{\omega_{P\alpha}^2}{\omega^2 +i\Gamma_{\alpha}\omega -
\omega_{\alpha}^2} \Big{)}
\end{equation}

Here $\Gamma_{\alpha}$ is the linewidth of the intersubband
plasmon modes, that can be determined in a phenomenological way,
i.e. from absorption measurements \cite{Book_Helm}.

On the basis on the results from the previous section we can
obtain an approximation for the dielectric constant, by replacing
the effective plasma thickness $L_{\mathrm{eff}}^\alpha$ with its
maximum value $L_{\mathrm{QW}}/f_{\alpha}^o$ (see equation
\pref{Leff_inequality}). We can call this substitution
"semiclassical approximation", since it leads to a plasma
self-energy provided by the classical expression, as seen from
equation \pref{omega_P_inequality}. This leads to the widespread
expression for the heterostructure dielectric constant
\cite{Wendler_Kandler_1993}:

\begin{eqnarray}
\varepsilon_\mathrm{QW, cl}(\omega) = \varepsilon \Big{(}
1-\frac{\bar{\omega}_{P\alpha}^2}{\omega^2 +i\Gamma_{\alpha}\omega
- \omega_{\alpha}^2} \Big{)}
\\
\bar{\omega}_{P\alpha}^2 = \frac{f_{\alpha}^o e^2 \Delta
N_{\alpha}}{\varepsilon \varepsilon_0 m^\ast S L_{\mathrm{QW}}}
\end{eqnarray}

In this limit, we obtain exactly the effective medium constant
from Ref.\cite{Zaluzny_Nalewajko_1999}, since in this case the
product $f_{\alpha}^o f_{\alpha}^w$ is replaced with
$L_{\mathrm{QW}}/L_{\mathrm{cav}}$, which is independent from the
particular intersubband transition. From section \ref{Sec_NoGo} it
is clear that the semiclassical approximation assumes that the
intersubband polarization is constant inside the heterostructure
slab, and zero everywhere else.

Including more than one transition in the definition of the
dielectric constant is not not trivial, since one is obliged to
diagonalize the full plasma Hamiltonian \pref{Hplasma_q} taking
into account the coupling between the different intersubband
plasmons. The semiclassical approximation allows to write a simple
analytical expression for the mutli-transition dielectric
function:

\begin{equation}\label{RPA_epQW}
\varepsilon_{\mathrm{QW, cl}}(\omega) = \varepsilon \Big{(} 1 -
\sum_{\alpha} \frac{\bar{\omega}^2_{P\alpha}}{\omega^2
+i\Gamma_{\alpha}\omega_\mathbf{q} - \omega_{\alpha}^2}\Big{)}
\end{equation}

This result is demonstrated in Appendix \ref{AppSCHplasma}.

\section{Conclusion}

In summary, we have provided a theoretical description of the
light-matter interaction for the intersubband transitions
in the electrical dipole gauge. We showed that, by
introducing a microscopic expression for the intersubband
polarization field $\mathbf{\hat{P}}(\mathbf{r})$, the
Power-Zienau-Woolley Hamiltonian provides a suitable framework for
the study of the interaction between the collective electronic
excitations and light. This description is very general and
applies to an arbitrary electronic potential, once the
corresponding single particle wavefunctions are known. In
particular, it can be used in the case of the ultra-strong
coupling regime, in the limit of very high electronic densities
and high overlap factors between the quantum well medium and the
microcavity mode.

The physical interpretation of the dipolar interaction Hamiltonian
is straightforward. The linear part of the Hamiltonian,
$\hat{H}_{I1}$ describes the coupling of the electronic
excitations with the cavity electromagnetic field. The quadratic
part, $\hat{H}_{I2}$, contains the depolarization effect (the
effect of an oscillating current on itself) and the coupling
between plasmons from different transitions.  The plasma energy
described by the term $\hat{H}_{I2}$ leads to the No-go theorem
for intersubband transitions. This energy has a close analogy the
electromagnetic dipole self-energy of a two level system
\cite{Book_Cohen_Ph_At}, that plays an important role for the
estimation of the radiative corrections of atomic transitions
\cite{Milonni_1976, Babiker_Loudon_1983}.

When the dipolar Hamiltonian is transformed back in the Coulomb
gauge, it provides the Coulomb interaction terms of the system
including image charge effects arising from the boundary
conditions on the cavity walls. This has been discussed for a
truncated Hamiltonian describing a single intersubband transition
interacting with a $\mathrm{TM}_0$ mode of a double metal
waveguide. In this case we find that the weights of the scalar
potential $V$ and the $\mathbf{A}^2$ term of the Coulomb gauge
version of the Hamiltonian are governed by the geometrical overlap
factor $f^w_\alpha$. Our study completes the theoretical framework
of the ultra-strong coupling, introduced in Ref.
\cite{Ciuti_PhysRevB_2005}, where only on the vector potential
part $\mathbf{A}^2$ was considered. We found that this study  is
justified for cavities with filling factors close to unity. A more
general PZW back transformation, which uses the whole basis of
guided modes and the full set of intersubband transitions will be
discussed elsewhere.

We believe that the approach developed here, based on operator
formalism to describe the collective intersubband excitations,
provides a compact and useful description that could allow further
studies of quantum electro-dynamical effects inserted in solid
state systems.

The authors acknowledge very useful discussions with C. Ciuti, D.
Hagenmuller and S. De Liberato. We also acknowledge financial
support from the ERC grant "ADEQUATE".

\appendix

\section{Vacuum field amplitude of the guided modes}\label{AppNormEMfield}

In this Appendix, we derive the vacuum normalization constant
$A_\mathbf{q}$ (equation \pref{A_q}) by taking into account the
multilayered geometry of the guided mode stack. We also comment on
the special treatment of metallic boundaries of the guiding
structure.

We start by expressing the volume integral \pref{H_p0} as a
function of the TM polarized field components (4)-(6):

\begin{eqnarray}
\frac{\mu_0}{2}\int \mathbf{\hat{H}}^2 d^3 \mathbf{r} =
\frac{\mu_0}{2}\int \mathbf{\hat{H}}\mathbf{\hat{H}^\dagger} d^3
\mathbf{r} =
\nonumber \\
S\sum_{\mathbf{q}} A_\mathbf{q}^2
(a^\dagger_\mathbf{q}a_\mathbf{q}+a_\mathbf{q}a^\dagger_\mathbf{q}+
a_\mathbf{q}a_\mathbf{-q}+a^\dagger_\mathbf{q}a^\dagger_\mathbf{-q})\times
\nonumber \\
\frac{\mu_0}{2} \int f_\mathbf{q}^2 (z)dz
\end{eqnarray}

\begin{eqnarray}
\int \frac{1}{2\varepsilon_0\varepsilon (z)}\mathbf{\hat{D}}_z^2
d^3 \mathbf{r} = \int \frac{1}{2\varepsilon_0\varepsilon (z)}
\mathbf{\hat{D}}_z\mathbf{\hat{D}}^\dagger_z d^3 \mathbf{r} =
\nonumber \\
S\sum_{\mathbf{q}} A_\mathbf{q}^2
(a^\dagger_\mathbf{q}a_\mathbf{q}+a_\mathbf{q}a^\dagger_\mathbf{q}-
a_\mathbf{q}a_\mathbf{-q}-a^\dagger_\mathbf{q}a^\dagger_\mathbf{-q})\times
\nonumber \\
 \frac{\mathbf{q}^2}{\omega_{c\mathbf{q}}^2}\int \frac{1}{2\varepsilon_0\varepsilon (z)}f_\mathbf{q}^2(z)dz
\end{eqnarray}

\begin{eqnarray}
\int \frac{1}{2\varepsilon_0\varepsilon
(z)}\mathbf{\hat{D}}_{\parallel}^2 d^3 \mathbf{r} = \int
\frac{1}{2\varepsilon_0\varepsilon (z)}
\mathbf{\hat{D}}_{\parallel}\mathbf{\hat{D}}^\dagger_{\parallel}
d^3 \mathbf{r} =
\nonumber \\
S\sum_{\mathbf{q}} A_\mathbf{q}^2
(a^\dagger_\mathbf{q}a_\mathbf{q}+a_\mathbf{q}a^\dagger_\mathbf{q}-
a_\mathbf{q}a_\mathbf{-q}-a^\dagger_\mathbf{q}a^\dagger_\mathbf{-q})\times
\nonumber \\
\frac{1}{\omega_{c\mathbf{q}}^2}\int
\frac{1}{2\varepsilon_0\varepsilon
(z)}\Big{(}\frac{df_\mathbf{q}}{dz}\Big{)}^2dz
\end{eqnarray}

To derive these expressions we have used the parity of
$\omega_{c\mathbf{q}}$ and $A_\mathbf{q}$ as respect to
$\mathbf{q}$ and
$\mathbf{e}_{-\mathbf{q}}=-\mathbf{e}_{\mathbf{q}}$. We have also
used the orthogonality of the two dimensional space harmonics
$\exp{(i\mathbf{q}\mathbf{r}_{\|})}/\sqrt{S}$.

To obtain the proper normalization of the electromagnetic field,
we first show the following identity:

\begin{equation}\label{A1_id}
I = \int dz \Big{[} \frac{\mu_0}{2}
f_\mathbf{q}^2-\frac{1}{2\varepsilon_0\varepsilon
(z)}\Big{(}\frac{\mathbf{q}^2}{\omega_{c\mathbf{q}}^2}f_\mathbf{q}^2+\Big{(}\frac{df_\mathbf{q}}{dz}\Big{)}^2\Big{)}
\Big{]}=0
\end{equation}

For the demonstration, we use the Helmholtz equation
\pref{f_q_equation}. Let us consider the $i^{\mathrm{th}}$ layer,
then by definition $\varepsilon(z)=\varepsilon_i$ is constant. By
multiplying the equation \pref{f_q_equation} by $f_\mathbf{q}$ and
integrating across the $i^{\mathrm{th}}$ layer we obtain:

\begin{eqnarray}
I_i = -
\frac{1}{\varepsilon_0\varepsilon_i}f_\mathbf{q}\frac{df_\mathbf{q}}{dz}\Big{|}_{i+}
+\frac{1}{\varepsilon_0\varepsilon_i}f_\mathbf{q}\frac{df_\mathbf{q}}{dz}\Big{|}_{i-}
\\
I_i =\int_i dz \Big{[} \mu_0
f_\mathbf{q}^2-\frac{1}{\varepsilon_0\varepsilon_i}
\Big{(}\frac{\mathbf{q}^2}{\omega_{c\mathbf{q}}^2}f_\mathbf{q}^2+\Big{(}\frac{df_\mathbf{q}}{dz}\Big{)}^2\Big{)}
\Big{]}
\end{eqnarray}

Here we have also used that $\varepsilon_0 \mu_0=1/c^2$, and the
symbol $i\pm$ means the upper/lower side of the layer. Using the
boundary conditions of the electromagnetic field:

\begin{eqnarray}
f_\mathbf{q}|_{i-} =f_\mathbf{q}|_{(i+1)+}
\\
\frac{1}{\varepsilon_i}\frac{df_\mathbf{q}}{dz}\Big{|}_{i-}
=\frac{1}{\varepsilon_{i+1}}\frac{df_\mathbf{q}}{dz}\Big{|}_{(i+1)+}
\end{eqnarray}

and the fact that $I = 1/2\sum_i I_i$ we finally obtain:

\begin{equation}
I =-
\frac{1}{2\varepsilon_0\varepsilon_U}f_\mathbf{q}\frac{df_\mathbf{q}}{dz}\Big{|}_{U}
+\frac{1}{2\varepsilon_0\varepsilon_D}f_\mathbf{q}\frac{df_\mathbf{q}}{dz}\Big{|}_{D}
\end{equation}

where $U/D$ denote the upper/lower boundary of the multi-layer.
Either one or the two boundaries go to infinity; then the field
and its derivative go to zero. The other option is to take
metallic boundaries. Since in the mid- and far- infrared range the
dielectric constant of the metals $\varepsilon_M$ is very high, we
can adopt perfect metal boundary conditions, where
$|\varepsilon_M|\rightarrow \infty$. We can then neglect the
contribution of the field from the metallic layers in the integral
\pref{H_p0}. In all cases the identity \pref{A1_id} is proven.
Thanks to this identity, the anti-resonant terms in the photon
Hamiltonian cancel, and the remaining term is simplified to:

\begin{equation}
\sum_{\mathbf{q}} 2\mu_0 S
L_{\mathbf{q}}A_\mathbf{q}^2(a^\dagger_\mathbf{q}a_\mathbf{q}+1/2)
\end{equation}

Following standard textbooks \cite{Book_Cohen_Ph_At}, the
prefactor should correspond to the energy quantum of the guided
mode $\hbar \omega_{c\mathbf{q}}$, which leads to equation \pref{A_q}.

Although we used perfect metal boundary conditions for the
normalization of the vacuum field amplitude $A_{\mathbf{q}}$, we
can still describe the effects of the finite metal permittivity
$\varepsilon_M$ in our model. As an example, consider the current
experimental situation where the multilayer is constituted by a
surface plasmon waveguide, obtained by deposing a metal layer on a
semiconductor substrate. We suppose the metal-semiconductor
interface to be at $z=0$. In the semiconductor the mode profile is
described by an exponential function $f_\mathbf{q}(z) =
e^{\gamma_\mathbf{q}z}$ for $z<0$, with $\gamma_\mathbf{q}$ the
decay length of the surface plasmon mode into the semiconductor.
Let $\varepsilon$ be the dielectric constant of the semiconductor,
then we have \cite{Book_Raether}:

\begin{eqnarray}
\frac{\omega_{c \mathbf{q}}^2}{c^2} = \mathbf{q}^2
\Big{|}\frac{1}{\varepsilon}+\frac{1}{\varepsilon_M} \Big{|}
\\
\gamma_\mathbf{q} = \varepsilon \frac{\omega_{c
\mathbf{q}}}{c}\frac{1}{|\varepsilon+\varepsilon_M|}
\end{eqnarray}

Using the above prescription, we consider the field in the metal to vanish,
but we keep the expression of the finite dielectric function
$\varepsilon_M$ in the dispersion relations above. We then have
$L_\mathbf{q} = 1/2\gamma_\mathbf{q}$, and the quantum amplitude
of the surface plasmon mode becomes:

\begin{equation}
A_\mathbf{q} =\Big{(} \frac{\varepsilon_0 \varepsilon c \hbar
\omega_{c \mathbf{q}}^2}{S |\varepsilon+\varepsilon_M|^{1/2}}
\Big{)}^{1/2}
\end{equation}

We see that in the far infrared domain, where $\omega_{c
\mathbf{q}} \rightarrow 0$ and $\varepsilon_M$ is very large and
negative, both spontaneous emission and the strong coupling
phenomena are unfavoured, since $A_\mathbf{q}$ has vanishingly
small values. On the contrary, the photon electric amplitude
$A_\mathbf{q}$ can be enhanced in a microcavity resonator.

\section{Derivation of the polartion dot Hamiltonian}\label{AppTransfo}

We first write the linear part of the interaction Hamiltonian
\pref{H_int} by using the expressions \pref{P_density}, \pref{D_m}
and the definition \pref{Fourier_trans}:

\begin{eqnarray}
H_{I1} = i\sum_{\alpha , m} \sqrt{\frac{\hbar
\omega_{cm}}{2\varepsilon \varepsilon_0 S
L_{\mathrm{cav}}}}z_\alpha(a^\dagger_m-a_m)\times
 \nonumber \\ \sum_{\mathbf{q}}[B^\dagger_{\alpha \mathbf{q}} \langle \mathbf{q}|u_m\rangle +
 B_{\alpha \mathbf{q}} \langle -\mathbf{q}|u_m\rangle]
\end{eqnarray}

Since the function $u_m(\mathbf{r}_{\parallel})$ is real, we have
$\langle -\mathbf{q}| u_m \rangle = \langle u_m| \mathbf{q}
\rangle$, and therefore we can use the transformation law
\pref{B_m_transf} and its hermitian conjugate to obtain:

\begin{eqnarray}
H_{I1} = i\sum_{\alpha , m} \sqrt{\frac{\hbar
\omega_{cm}}{2\varepsilon \varepsilon_0 S
L_{\mathrm{cav}}}}z_\alpha(a^\dagger_m-a_m)\times
 \nonumber \\  (B^\dagger_{\alpha m} + B_{\alpha m})
\end{eqnarray}

In order to transform the quadratic part \pref{H_I2} we use the
inverse transformations:

\begin{eqnarray}
B^\dagger_{\alpha \mathbf{q}} = \sum_m \langle u_m|
\mathbf{q}\rangle B^\dagger_{\alpha m}
\\
B_{\alpha \mathbf{q}} = \sum_m \langle \mathbf{q}|u_m\rangle
B_{\alpha m}
\end{eqnarray}

and therefore:

\begin{eqnarray}\label{Bsquare_m}
\sum_\mathbf{q} (B^\dagger_{\alpha \mathbf{q}}+B_{\alpha
\mathbf{-q}})(B^\dagger_{\beta \mathbf{-q}}+B_{\beta \mathbf{q}}) = \nonumber \\
\sum_{\mathbf{q}, m, m'} (\langle u_m| \mathbf{q}\rangle
B^\dagger_{\alpha m} + \langle \mathbf{-q}|u_m\rangle B_{\alpha
m})\times
 \nonumber \\ (\langle u_{m'}| \mathbf{-q}\rangle
B^\dagger_{\beta {m'}} + \langle \mathbf{q}|u_{m'}\rangle B_{\beta
{m'}})
\end{eqnarray}

Using once again the relation $\langle -\mathbf{q}| u_m \rangle =
\langle u_m| \mathbf{q} \rangle$ and the closure relation
$\sum_\mathbf{q}|\mathbf{q}\rangle \langle \mathbf{q}|=1$ along
with the orthogonality condition $\langle  u_m| u_{m'} \rangle
=\delta_{mm'}$ it is straightforward to transform the expression
\pref{Bsquare_m} into:

\begin{equation}
\sum_m (B^\dagger_{\alpha m} + B_{\alpha m})(B^\dagger_{\beta m} +
B_{\beta m})
\end{equation}

We then use exactly the same bosonization approach as in section
\ref{Bright_and_Dark} to arrive at the Hamiltonian
\pref{Hplasma_m}.

\section{Long wavelength limit of the Coulomb interaction}\label{AppCoulomb}

In order to show that the last term of equation \pref{H_pA} has the
form of the Coulomb interaction, we first re-express it through the
$B$-operators defined in section \ref{Sec_MicroDefP}:

\begin{eqnarray}\label{Vterm}
\sum_{\mathbf{q}} \frac{\hbar
\omega_P^2}{4\omega_{21}}(b^\dagger_{\mathbf{q}}+b_{-\mathbf{q}})(b^\dagger_{-\mathbf{q}}+b_{\mathbf{q}})
=\frac{1}{2S\varepsilon \varepsilon_0}\times \nonumber \\
\frac{e^2 \hbar^2}{4{m^\ast}^2 \omega_{21}^2}I_{12,12}
\sum_\mathbf{q}
(B^\dagger_{21\mathbf{q}}+B_{21-\mathbf{q}})(B^\dagger_{21-\mathbf{q}}+B_{21\mathbf{q}})
\end{eqnarray}

We first use the Vinter's indenity \cite{Ando_Fowler_Stern_1982,
Vinter_PRB_1977}:

\begin{eqnarray}
\frac{\hbar^2}{4{m^\ast}^2 \omega_{21}^2}I_{12,12} = \nonumber
\\ -\iint_{-\infty}^{+\infty} dz dz'
\phi_1 (z)\phi_2 (z) |z-z'| \phi_1 (z')\phi_2 (z')
\end{eqnarray}

and we recognize the prefactor of \pref{Vterm} to be exactly the
long wavelength limit of the Coulomb interaction matrix element
$V^{\lambda \mu, \mu' \lambda'}_{\mathbf{q}\rightarrow
\mathbf{0}}$:

\begin{eqnarray}
V^{\lambda \mu, \mu' \lambda'}_{\mathbf{q}\rightarrow \mathbf{0}}
= -\frac{e^2}{2S\varepsilon \varepsilon_0}\times  \nonumber
\\   \iint_{-\infty}^{+\infty} dz dz' \phi_\lambda (z)\phi_\mu (z)
|z-z'| \phi_{\mu'} (z')\phi_{\lambda'} (z')
\end{eqnarray}

Note that we have $V^{1 2, 1 2}_{\mathbf{q}\rightarrow \mathbf{0}}
=V^{2 1, 2 1}_{\mathbf{q}\rightarrow \mathbf{0}}=V^{2 1, 1
2}_{\mathbf{q}\rightarrow \mathbf{0}}=V^{1 2, 2
1}_{\mathbf{q}\rightarrow \mathbf{0}}$, so we recover only the
matrix elements that describe the interaction of electrons between
subbands 1 and 2, as should be expected.

Next, using the fermionic commutation rules, we can easily
rearrange the binary products of $B$-operators into four
$c$-operator products:

\begin{equation}
\sum_\mathbf{q} B^\dagger_{21\mathbf{q}}B^\dagger_{21-\mathbf{q}}
= \sum_{\mathbf{q}, \mathbf{k}, \mathbf{k}'}c^\dagger_{2
\mathbf{k+q}}c^\dagger_{2 \mathbf{k'-q}}c_{1 \mathbf{k'}}c_{1
\mathbf{k}}
\end{equation}

\begin{equation}
\sum_\mathbf{q} B_{21\mathbf{q}}B_{21-\mathbf{q}} =
\sum_{\mathbf{q}, \mathbf{k}, \mathbf{k}'}c^\dagger_{1
\mathbf{k+q}}c^\dagger_{1 \mathbf{k'-q}}c_{2 \mathbf{k'}}c_{2
\mathbf{k}}
\end{equation}

\begin{eqnarray}\label{B^+B}
\sum_\mathbf{q} B^\dagger_{21\mathbf{q}}B_{21\mathbf{q}} =
\sum_{\mathbf{q}, \mathbf{k}, \mathbf{k}'}c^\dagger_{2
\mathbf{k+q}}c^\dagger_{1 \mathbf{k'-q}}c_{2 \mathbf{k'}}c_{1
\mathbf{k}} \nonumber
\\  + \sum_{\mathbf{q}, \mathbf{k}}c^\dagger_{2
\mathbf{k+q}}c_{2 \mathbf{k+q}}
\end{eqnarray}

\begin{eqnarray}\label{BB^+}
\sum_\mathbf{q} B_{21\mathbf{q}}B^\dagger_{21\mathbf{q}} =
\sum_{\mathbf{q}, \mathbf{k}, \mathbf{k}'}c^\dagger_{1
\mathbf{k+q}}c^\dagger_{2 \mathbf{k'-q}}c_{1 \mathbf{k'}}c_{2
\mathbf{k}} \nonumber
\\  + \sum_{\mathbf{q}, \mathbf{k}}c^\dagger_{1
\mathbf{k+q}}c_{1 \mathbf{k+q}}
\end{eqnarray}

Note that the sum of the pair terms in \pref{B^+B} and \pref{BB^+}
simply acts as the identity operator in the two subband subspace
and therefore can be ignored. The remaining four operator terms
can be regrouped in order to provide the long wavelength expansion
of the Coulomb potential $\bar{V}$ for our problem:

\begin{eqnarray}\label{V_barre}
\bar{V} = (1-f_{21}^o f_{21}^w)\times \nonumber \\
\sum_{\substack {\mathbf{q}, \mathbf{k}, \mathbf{k}' \\ [\lambda
\mu, \mu' \lambda'] }} V^{\lambda \mu, \mu' \lambda'
}_{\mathbf{q}\rightarrow \mathbf{0}}c^\dagger_{\lambda
\mathbf{k+q}}c^\dagger_{\mu \mathbf{k'-q}}c_{\mu'
\mathbf{k'}}c_{\lambda' \mathbf{k}}
\end{eqnarray}

The symbol $[\lambda \mu, \mu' \lambda']$ means that the sum runs
only along the four Coulomb matrix element mentioned above.

The final result \pref{V_barre} can be interpreted as a
dipole-dipole interaction, where the oscillating intersubband
dipole moments interact with each other through their local field
\cite{Hopfield_1958}.

\section{Properties of the polariton dispersion}\label{AppPropDisp}

The eigenvalue equation for the polariton problem is (equation
\pref{EigenEq1}):

\begin{equation}\label{Eigen2}
(\omega^2 - \widetilde{\omega}_{\alpha}^2)(\omega^2 - \omega_c^2)
= f_{\alpha}^o f_{\alpha}^w \omega_P^2 \omega_c^2
\end{equation}

For simplicity we have dropped the wavevector index $\mathbf{q}$.
The real solution of equation \pref{Eigen2} are obtained from:

\begin{equation}\label{Solutions}
\omega^2_{\pm} =
\frac{1}{2}(\omega_c^2+\widetilde{\omega}_{\alpha}^2 \pm
\sqrt{\Delta})
\end{equation}

\begin{equation}\label{DeltaEq}
\Delta =
(\omega_c^2+\widetilde{\omega}_{\alpha}^2)^2-4\omega_c^2\bar{\omega}_{\alpha}^2
\end{equation}

From these equations it is easy to deduce:

\begin{equation}\label{Usefull_1}
\frac{\mathrm{d}\omega_{\pm}}{\mathrm{d}\omega_c^2}
=\frac{1}{4\omega_{\pm}}\Big{(} 1 \pm
\frac{\mathrm{d}\sqrt{\Delta}}{\mathrm{d}\omega_c^2} \Big{)}
\end{equation}

The minimal splitting is obtained from:

\begin{equation}
\frac{\mathrm{d}(\omega_{+}-\omega_{-})}{\mathrm{d}\omega_c^2}=0
\end{equation}

Multiplying this equation by $\omega_{+}-\omega_{-}$ and using
\pref{Usefull_1} we obtain:

\begin{equation}
(\omega_{+}-\omega_{-})^2 =
\frac{\omega_{+}^2-\omega_{-}^2}{2\sqrt{\Delta}}\frac{\mathrm{d}\sqrt{\Delta}}{\mathrm{d}\omega_c^2}
\end{equation}

To transform this equation, we use the following relations:

\begin{eqnarray}
\omega_{+}^2-\omega_{-}^2 =\sqrt{\Delta} \\
\omega_{+}^2+\omega_{-}^2
=\omega_c^2+\widetilde{\omega}_{\alpha}^2 \\
\omega_{+}^2\omega_{-}^2 =\omega_c^2\bar{\omega}_{\alpha}^2
\end{eqnarray}

The first is an immediate corollary from \pref{Solutions}, while
the second and the third are the Newton formulas for equation
\pref{Eigen2}. We then have:

\begin{eqnarray}
2(\omega_{+}-\omega_{-})^2 =2(\omega_{+}^2+\omega_{-}^2 -
2\omega_{+}\omega_{-}) \nonumber \\
=2(\omega_c^2+\widetilde{\omega}_{\alpha}^2 -
2\omega_c\bar{\omega}_{\alpha})
=\frac{\mathrm{d}\sqrt{\Delta}}{\mathrm{d}\omega_c^2} \nonumber \\
= 2(\omega_c^2+\widetilde{\omega}_{\alpha}^2) -
4\bar{\omega}_{\alpha}^2
\end{eqnarray}

To obtain the last line, we have derived equation
\pref{DeltaEq} with respect to $\omega_c^2$. This equation clearly
leads to the result:

\begin{equation}
\frac{\mathrm{d}(\omega_{+}-\omega_{-})}{\mathrm{d}\omega_c^2}=0
\phantom{Q}\Leftrightarrow
\phantom{Q}\omega_c=\bar{\omega}_{\alpha}
\end{equation}

Furthermore, we can express the minimal splitting:

\begin{equation}
\mathrm{min} (\omega_{+}-\omega_{-})^2
=\widetilde{\omega}_{\alpha}^2 -\bar{\omega}_{\alpha}^2
=f_{\alpha}^o f_{\alpha}^w \omega_P^2
\end{equation}

This is the result stated in equations \pref{Rabi1} and
\pref{Rabi2}.

\section{Semiclassical plasma Hamiltonian}\label{AppSCHplasma}

In this Appendix we consider the plasma Hamiltonian in the
semi-classical approximation discussed in the end of section
\ref{DielFuncSection}. It will be shown that this Hamiltonian
leads to the semi-classical dielectric constant \pref{RPA_epQW}.

In the semi-classical approximation, it is assumed that the
intersubband polarization is constant along the heterostructure
slab. We can then approximate the current-current correlation
function by the first order in the expansion \pref{I_expansion}
described in section \ref{Sec_NoGo}:

\begin{eqnarray}\label{I_approx}
\int_{-\infty}^{+\infty} \xi_{\alpha} (z) \xi_{\beta} (z) dz \approx \nonumber \\
\frac{1}{L_{\mathrm{QW}}}\int_0^{L_{\mathrm{QW}}} \xi_{\alpha} (z)
dz \int_0^{L_{\mathrm{QW}}} \xi_{\beta} (z) dz = \nonumber \\
\frac{2 m^\ast}{\hbar}\sqrt{\omega_\alpha \omega_\beta f_\beta^o
f_\alpha^o}
\end{eqnarray}

This is equivalent to use a suitably averaged microscopic response
in the quantum well slab. With the use of \pref{I_approx}, we
obtain the following expressions for the coefficients that enter
the plasma Hamiltonian:

\begin{eqnarray}
L^{\alpha}_{\mathrm{eff}} = L_{\mathrm{QW}}/f_\alpha^o
\\
\bar{\omega}^2_{P_\alpha} = \frac{e^2 f_\alpha^o \Delta
N_\alpha}{\varepsilon \varepsilon_0 m^\ast S L_{\mathrm{QW}}}
\\
C_{\alpha, \beta} = 1
\end{eqnarray}

We consider first the matter part of the plasma Hamiltonian, which
is rewritten as:

\begin{eqnarray}\label{AppCHplasma_q}
H =  \sum_{\alpha, \mathbf{q}}\hbar \widetilde{\omega}_{\alpha}
p^\dagger_{\alpha
\mathbf{q}}p_{\alpha \mathbf{q}}+ \nonumber\\
\sum_{\alpha \neq \beta, \mathbf{q}} \frac{\hbar
\bar{\omega}_{P\alpha} \bar{\omega}_{P\beta}}{2
\sqrt{\widetilde{\omega}_{\alpha}
\widetilde{\omega}_{\beta}}} \times \nonumber\\
(p^\dagger_{\alpha \mathbf{q}}+p_{\alpha
-\mathbf{q}})(p^\dagger_{\beta -\mathbf{q}}+p_{\beta \mathbf{q}})
\end{eqnarray}

Since the plasmon coupling coefficients are independent from the
the index $\mathbf{q}$ the latter is dropped in the equations, and
\pref{AppCHplasma_q} is expressed in a more handy form:

\begin{eqnarray}
H =  \sum_{\alpha}\hbar \widetilde{\omega}_{\alpha}
p^\dagger_{\alpha}p_{\alpha}+ \sum_{\alpha \neq \beta} \hbar
\Omega_{\alpha \beta}
(p^\dagger_{\alpha}+p_{\alpha})(p^\dagger_{\beta}+p_{\beta})\phantom{QQ}
\end{eqnarray}

\begin{equation}\label{AppCdefOm}
\Omega_{\alpha \beta} =\Omega_{\beta \alpha} = \frac{\hbar
\bar{\omega}_{P \alpha} \bar{\omega}_{P \beta}}{2
\sqrt{\widetilde{\omega}_\alpha \widetilde{\omega}_\beta}}
\end{equation}

Let us consider $N$ coupled plasmons; then the Hopfield
determinant of the resulting Hamiltonian is:

\begin{equation}
\left | \begin{array}{ccccccc}
\omega - \widetilde{\omega}_1 & 0 & -\Omega_{12} & -\Omega_{12} & \cdots &-\Omega_{1N} & -\Omega_{1N}\\
0 & \omega + \widetilde{\omega}_1 & \Omega_{12} & \Omega_{12}  & \cdots & \Omega_{1N} &  \Omega_{1N}\\
-\Omega_{12} & -\Omega_{12} & \omega - \widetilde{\omega}_2 & 0 & \cdots &-\Omega_{2N} & -\Omega_{2N} \\
\Omega_{12} & \Omega_{12} & 0 & \omega + \widetilde{\omega}_2 &
\cdots & \Omega_{2N} & \Omega_{2N} \\
\vdots & \vdots & \vdots & \vdots & \ddots & \vdots & \vdots \\
-\Omega_{1N} & -\Omega_{1N} & -\Omega_{2N} & -\Omega_{2N} & \cdots
& \omega - \widetilde{\omega}_N & 0 \\
\Omega_{1N} & \Omega_{1N} & \Omega_{2N} & \Omega_{2N} & \cdots & 0
& \omega + \widetilde{\omega}_N
\end{array}\right |
\end{equation}

To simplify this determinant, we first add every impair row to the
row above, then we substract every impair column from the column
on the left. This leads to the following simplified determinant:

\begin{equation}
\left | \begin{array}{ccccccc}
\omega - \widetilde{\omega}_1 & 2\widetilde{\omega}_1 & 0 & 0 & \cdots & 0 & 0 \\
0 & \omega + \widetilde{\omega}_1 & \Omega_{12} & 0  & \cdots & \Omega_{1N} &  0\\
0 & 0 & \omega - \widetilde{\omega}_2 & 2\widetilde{\omega}_2 & \cdots & 0 & 0 \\
\Omega_{12} & 0 & 0 & \omega + \widetilde{\omega}_2 & \cdots & \Omega_{2N} & 0 \\
\vdots & \vdots & \vdots & \vdots & \ddots & \vdots & \vdots \\
0 & 0 & 0 & 0 & \cdots & \omega - \widetilde{\omega}_N & 2\widetilde{\omega}_N \\
\Omega_{1N} & 0 & \Omega_{2N} & 0 & \cdots & 0 & \omega +
\widetilde{\omega}_N
\end{array}\right |
\end{equation}

Let $\Delta_{[1...N]}$ be the determinant of $N$-th order.
Developing this determinant along the columns we obtain the
following recursive expression:

\begin{eqnarray}
\Delta_{[1...N]} = (\omega^2 -
\widetilde{\omega}_1^2)\Delta_{[2...N]}
\nonumber\\
- 2^2\sum_{i}^N \widetilde{\omega}_1\widetilde{\omega}_i
\Omega_{1i}\Omega_{i1}\Delta_{[2...N]/[i]} \nonumber\\
...
\nonumber\\
-n!2^{n+1}\sum_{\overline{i_1, i_2...i_n}}\widetilde{\omega}_1
\widetilde{\omega}_{i_1}...\widetilde{\omega}_{i_n}\Omega_{1i_1}\Omega_{i_1i_2}
...\Omega_{i_n 1}\times
\nonumber\\
\Delta_{[2...N]/[i_1, i_2...i_n]} \nonumber\\...
\nonumber\\
-(N-1)!
2^N\widetilde{\omega}_1\widetilde{\omega}_2...\widetilde{\omega}_N\Omega_{12}\Omega_{23}...\Omega_{N1}\phantom{QQ}
\end{eqnarray}

with $\Delta_{[2...N]/[i_1, i_2...i_n]}$ being the determinant of
order $N-1-n$, excluding the transitions $i_1, i_2...i_n$. The
notation $\overline{i_1, i_2...i_n}$ means that the sum does not
contain repetitive indexes. Using the definition \pref{AppCdefOm}
the determinant is easily rewritten as:

\begin{eqnarray}
\Delta_{[1...N]} = (\omega^2 -
\widetilde{\omega}_1^2)\Delta_{[2...N]} - \sum_{i}^N
\bar{\omega}_{P1}^2\bar{\omega}_{Pi}^2\Delta_{[2...N]/[i]} \nonumber\\
...
\nonumber\\
-n!\sum_{\overline{i_1, i_2...i_n}}\bar{\omega}_{P1}^2
\bar{\omega}_{Pi_1}^2...\bar{\omega}_{Pi_n}^2\Delta_{[2...N]/[i_1,
i_2...i_n]} \nonumber\\...
\nonumber\\
-(N-1)!\bar{\omega}_{P1}^2\bar{\omega}_{P2}^2...\bar{\omega}_{PN}^2\phantom{QQ}
\end{eqnarray}

Having expressed the $N$-th order determinant as a function of
lower order ones, we make the following recursive assumption for
$n<N$:

\begin{equation}
\Delta_{[1...n]} =
\prod_{i=1}^n(\omega^2-\omega_i^2)\Big{(}1-\sum_{i=1}^n\frac{\bar{\omega}_{Pi}^2}{\omega^2-\omega_i^2}\Big{)}
\end{equation}

Let us denote $x_i = \bar{\omega}_{Pi}^2/(\omega^2-\omega_i^2)$,
then our recursive hypothesis implies:

\begin{eqnarray}
\Delta_{[1...N]} = \prod_{i=1}^N(\omega^2-\omega_i^2) \times
\nonumber \\ \Big{[} (1-x_1)\Big{(}1-\sum_{i=2}^N x_i \Big{)} -
\sum_{i=2}^N x_1 x_i \Big{(}1-\sum_{k=2,k\neq i}^N x_k \Big{)}
\nonumber \\
...-n!\sum_{\overline{i_1, i_2...i_n}} x_1x_{i_1}...x_{i_n}
\Big{(} 1-\sum_{k=2,k\neq i_1,i_2...i_n}^N
x_k \Big{)} \nonumber \\
...-(N-1)!x_1 x_2 ... x_N \Big{]}\phantom{QQ}
\end{eqnarray}

We can easily show that all the terms in this of sum, that imply
high orders of the $x_i$ cancel two by two, except the linear
terms, which completes our recursive demonstration. The final
result can be cast in the form:

\begin{equation}\label{Det_Annex}
\Delta_{[1...N]} = \frac{\varepsilon_{\mathrm{QW,
cl}}(\omega)}{\varepsilon}\prod_{i=1}^N(\omega^2-\omega_i^2)
\end{equation}

where $\varepsilon_{\mathrm{QW, cl}}(\omega)$ is the
semi-classical multi-band dielectric constant, provided by
equation \pref{RPA_epQW}. Therefore the coupled plasmonic
eigenmodes of the Hamiltonian \pref{AppSCHplasma} are provided by
the zeroes of the dielectric function
$\varepsilon_{\mathrm{QW, cl}}(\omega)$.

If we now add the light-matter interaction with a single photonic
mode, similar reasoning leads to the Hopfield determinant:

\begin{equation}
\Delta(\omega) = \frac{\varepsilon_{\mathrm{QW,
cl}}(\omega)}{\varepsilon}
\Big{(}\omega^2-\omega_c^2\frac{\varepsilon}{\varepsilon_{\mathrm{eff,
cl}}(\omega)} \Big{)}\prod_{i=1}^N(\omega^2-\omega_i^2)
\end{equation}

with $\varepsilon_{\mathrm{eff, cl}}(\omega)$ the classical effective medium
constant defined by:

\begin{equation}
\frac{1}{\varepsilon_{\mathrm{eff, cl}}(\omega)} =
\frac{f_w}{\varepsilon_{\mathrm{QW,
cl}}(\omega)}+\frac{1-f_w}{\varepsilon}
\end{equation}

Here $f_w = L_{\mathrm{QW}}/L_\mathrm{cav}$ is the filling factor
of the quantum well (heterostructure) in the cavity, which is
independent from the intersubband transition in this
approximation.

From a mathematical point of view, the analytical diagonalisation
of the problem was obtained thanks to the relation $C_{\alpha,
\beta}=1$. This would not be possible in the general case, and yet
it is clear that the Hopfield determinant of the Hamiltonian
\pref{Hplasma_q} is a polynomial of $2\times(N+1)$-th degree.
Therefore we can still use the definition \pref{deff_ep_eff} to
compute numerically the exact quantum effective dielectric
constant of the problem.

\end{document}